\documentclass[useAMS,usenatbib]{aa}
\usepackage{txfonts,color}
\usepackage{natbib}
\usepackage[T1]{fontenc} 
\usepackage{multirow}
\bibpunct{(}{)}{;}{a}{}{,}
\usepackage[dvips]{graphicx}

\newcommand{\simgt}{\lower.5ex\hbox{$\; \buildrel > \over \sim \;$}}
\newcommand{\simlt}{\lower.5ex\hbox{$\; \buildrel < \over \sim \;$}}

\newcommand{\hbe}{H\ensuremath{\beta}}
\newcommand{\hde}{H\ensuremath{\delta}}
\newcommand{\mgb}{\ion{Mg}{1}$b$}
\newcommand{\lam}[1]{\ensuremath{\lambda}#1}

\def\degree{$^\circ$}
\def\hh{$^{\mathrm h}$}
\def\mm{$^{\mathrm m}$}

\def\farcm{\hbox{$.\!\!^{\prime}$}}
\def\farcs{\hbox{$.\!\!^{\prime\prime}$}}

\def\fs{\hbox{$.\!\!{}^{\rm s}$}}
\def\ls{\mathrel{\hbox{\rlap{\hbox{\lower4pt\hbox{$\sim$}}}\hbox{$<$}}}}
\def\gs{\mathrel{\hbox{\rlap{\hbox{\lower4pt\hbox{$\sim$}}}\hbox{$>$}}}}

\begin{document}

\title{Probing cluster dynamics in RXCJ1504.1-0248 via radial and
  two-dimensional gas and galaxy properties}

\author{Yu-Ying Zhang\inst{1}, 
Miguel Verdugo\inst{2,3},
Matthias Klein\inst{1},
\and
Peter Schneider\inst{1}}

\institute{Argelander-Institut f\"ur Astronomie, Universit\"at Bonn,
  Auf dem H\"ugel 71, 53121 Bonn, Germany
  \and Max-Planck-Institut f\"ur extraterrestrische Physik,
  Giessenbachstra{\ss}e, 85748 Garching, Germany 
\and {University of Vienna, Department of Astronomy, 
T\"urkenschanzstra\ss e
17, 1180 Vienna, Austria}
}

\authorrunning{Zhang et al.}

\titlerunning{Probing cluster dynamics in RXCJ1504.1-0248 via radial and
  two-dimensional gas and galaxy properties}

\date{Received 07/02/2012 / Accepted 12/04/2012}

\offprints{Y.-Y. Zhang}

\abstract{We study one of the most X-ray luminous cluster of galaxies
  in the REFLEX survey, RXC J1504.1-0248 (hereafter R1504; $z_{\rm cl}
  = 0.2153$), using \emph{XMM-Newton} X-ray imaging spectroscopy,
  VLT/VIMOS optical spectroscopy, and WFI optical imaging. The mass
  distributions were determined using both the so-called hydrostatic
  method with X-ray imaging spectroscopy and the dynamical method with
  optical spectroscopy, respectively, which yield $M^{\rm
    H.E.}_{500}=(5.81\pm 0.49)\times 10^{14}M_{\odot}$ and $M^{\rm
    caustic}_{500}=(4.17\pm 0.42)\times 10^{14}M_{\odot}$. According
  to recent calibrations, the richness-derived mass estimates closely
  agree with the hydrostatic and dynamical mass estimates. The
  line-of-sight velocities of spectroscopic members reveal a group of
  galaxies with high velocities ($>$1000\,km\,s$^{-1}$) at a projected
  distance of about $r^{\rm H.E.}_{500}=(1.18\pm 0.03)$~Mpc south-east
  of the cluster centroid, which is also indicated in the X-ray
  two-dimensional (2-D) temperature, density, entropy, and pressure
  maps. The dynamical mass estimate is 80\% of the hydrostatic mass
  estimate at $r^{\rm H.E.}_{500}$. It can be partially explained by
  the $\sim$20\% scatter in the 2-D pressure map that can be
  propagated into the hydrostatic mass estimate. The uncertainty in
  the dynamical mass estimate caused by the substructure of the high
  velocity group is $\sim$14\%. The dynamical mass estimate using blue
  members is 1.23 times that obtained using red members. The global
  properties of R1504 obey the observed scaling relations of nearby
  clusters, although its stellar-mass fraction is rather low.
\begin{keywords}
  Cosmology: observations --- Galaxies: clusters: individual: RXCJ
  1504.1-0248 --- Galaxies: clusters: intracluster medium --- Methods:
  data analysis --- X-rays: galaxies: clusters --- Galaxy: kinematics
  and dynamics
\end{keywords}
}

\maketitle

\section{Introduction}

Simulations show that the formation of galaxy clusters is not a purely
gravitational process. For instance, the galaxy velocity dispersion in
clusters provides evidence of heating when compared to the cold dark
matter velocity dispersion normalized to the results from the
Wilkinson Microwave Anisotropy Probe and large-scale structure
distributions (e.g. Evrard et al.  2008). Cluster mergers do not only
disturb the X-ray appearance of the hot intracluster medium (ICM;
e.g. Ricker \& Sarazin 2001; Poole et al. 2006; Zhang et al. 2009;
B\"ohringer et al. 2010), but also affect the properties of the
cluster galaxies (e.g. Sun et al.  2007). The ICM and galaxies react
on different timescales during merging (e.g. Roettiger et
al. 1999). Clusters with similar X-ray global properties can contain
rather different galaxy populations (e.g. Popesso et al. 2007). There
is a higher fraction of blue galaxies in rich galaxy clusters at $z\gs
0.2$ than in local clusters (Butcher \& Oemler 1978, the so-called
Butcher-Oemler effect). This effect may be due to episodes of star
formation in a subset of cluster members driven by merging
(e.g. Saintonge et al. 2008). The kinematic analysis thus complements
the X-ray analysis in probing the dynamical structure of a cluster and
the consequences of this for the mass measurement.

Comparisons between X-ray and weak-lensing mass estimates suggest that
there is a 10--20\% deviations from hydrostatic equilibrium at
$r_{500}$ for relaxed clusters of galaxies, and yield controversial
findings for disturbed clusters based on different observations
(e.g. Zhang et al. 2008, 2010; Mahdavi et al. 2008). The comparison
between the X-ray hydrostatic and optical dynamical mass measurements
complements the X-ray versus weak-lensing mass comparison, and can
thus cross check deviations from hydrostatic equilibrium. Lensing is
strongly affected by projection effects because it is sensitive to all
mass along the line-of-sight. Line-of-sight velocities of cluster
galaxies via the redshifts of cluster galaxies can probe substructures
along the line-of-sight. Even for undisturbed clusters, the
dynamically relaxed central regions are surrounded by infalling
regions in which galaxies are bound to the clusters but not in
equilibrium (e.g. Biviano et al. 2006; Rines \& Diaferio
2006). Optical spectroscopy can constrain the dynamical properties out
to large radii (e.g. Braglia et al. 2007, Ziparo et al. 2012), where
it is difficult to probe with current X-ray satellites. On the other
hand, X-ray data can be used to derive two-dimensional (2-D) cluster
morphology projected on the sky. We use these two tools together with
the optical colour information to constrain substructures, and model
causes of deviations from hydrostatic equilibrium.

RXC J1504.1-0248 (hereafter R1504; $z_{\rm cl} = 0.2153$), a prominent
cool-core cluster, is one of the most relaxed clusters in the REFLEX
galaxy cluster survey (B\"ohringer et al. 2001). B\"ohringer et
al. (2005) investigated the central region of R1504 with
\emph{Chandra}, and found that the core region of R1504 appears to be
quite relaxed. Using optical data obtained from integral field
spectroscopy and X-ray data from \emph{XMM-Newton}, Ogrean et
al. (2010) found that feedback from active galactic nuclei (AGN) can
slow down the cooling flow to the observed mass deposition rate in
R1504, if the black-hole accretion rate is of the order of $0.5
M_{\odot}$~yr$^{-1}$ with 10\% energy output efficiency. The
\emph{XMM-Newton} data in the R1504 field cover about a two times
larger radial range than the \emph{Chandra} data, and the large
effective area of \emph{XMM-Newton} ensures high quality photon
statistics for a 2-D spectral analysis. Follow-up observations of
R1504 were also performed by the ESO Visible Multi-Object Spectrograph
at the 8.2\,m UT3 of the Very Large Telescope (VLT/VIMOS) out to about
$3r_{500}$. In addition, we have at our disposal optical imaging data
from the ESO Wide-Field Imager (WFI) in three bands and Subaru in two
bands.

We present our analysis of both the radial and 2-D gas and galaxy
properties of R1504 based on \emph{XMM-Newton}, VLT/VIMOS, and WFI
data as follows. We describe the X-ray analysis and results in
Sect.\,\ref{s:xray}, and optical spectroscopic and imaging analyses
and results in Sect.\,\ref{s:optspe} and Sect.\,\ref{s:optima}. We
discuss the systematic uncertainties in and interpretations of the
results in Sect.\,\ref{s:mcom}. In Sect.\,\ref{s:con}, we summarize
our conclusions. Throughout the paper, we assume that $\Omega_{\rm
  m}=0.3$, $\Omega_\Lambda=0.7$, and
$H_0=70$\,km\,s$^{-1}$\,Mpc$^{-1}$. Thus, an extent of 1\arcmin\ at
the distance of R1504 corresponds to 209.8\,kpc. Confidence intervals
correspond to the 68\% confidence level.

\section{X-ray data analysis and results}
\label{s:xray}

\subsection{{\it XMM-Newton} observations and data preparation}

The \emph{XMM-Newton} observations (ID: 0401040101) were taken in 2007
with a thin filter in full frame (FF) mode for MOS and extended full
frame (EFF) mode for pn. The fraction of out-of-time (OOT) effects
amounts to 2.32\%, which we used to normalize the pn OOT product
before subtracting it from the pn normal product.

We applied an iterative $2\sigma$ clipping procedure (e.g. Sect.\,2.1
of Zhang et al. 2006) to filter flares using the light curves in both
the soft band (0.3--10\,keV) binned in 10\,s intervals and the hard
band (10--12\,keV for MOS and 12--14\,keV for pn) binned in 100\,s
intervals. There are 34.9\,ks clean data for MOS1, 34.5\,ks for MOS2,
and 31.6\,ks for pn left. We generated a list of bright point sources
detected with the {\tt XMMSAS} task {\tt edetect\_chain} applied to
five energy bands: 0.3--0.5\,keV, 0.5--2\,keV, 2--4.5\,keV,
4.5--7.5\,keV, and 7.5--12\,keV. All confirmed sources were subtracted
from the event lists using a radius of 25\arcsec, which encloses
$\sim$80\% of the flux of the point sources according to the
\emph{XMM-Newton} point spread function (PSF). A weight column was
created with the {\tt XMMSAS} {\tt evigweight} command, which accounts
for the vignetting correction for off-axis observations.

\subsection{Temperature and surface brightness distributions}

\begin{figure*}
\begin{center}
\includegraphics[angle=270,width=8cm]{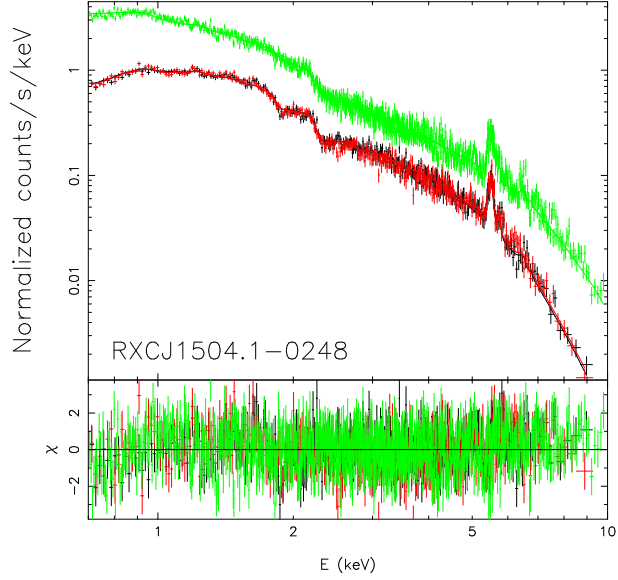}
\end{center}
\caption{MOS1 (black), MOS2 (red), and pn (green) spectra extracted from the
  central $\le $0\farcm5 circle. The lower panel shows deviations of the 
 models from the data points normalized by the $1\sigma$ errors.
 \label{f:spec}}
\end{figure*}

Following Sect.\,2.3 of Zhang et al. (2010), we measured the X-ray
flux-weighted centroid to be RA=15\hh04\mm07\fs801,
$\delta$=$-$02\degree48\arcmin10\farcs29 (J2000). We applied the
double-background subtraction method developed for clusters at similar
redshifts by Zhang et al. (2006, the so-called DBS II method
therein). We chose the \emph{XMM-Newton} blank sky accumulations in
the \emph{Chandra} Deep Field South (CDFS), which used the medium
filter and were also acquired in FF/EFF mode for MOS/pn as R1504, as
background observations. The count rate in the hard band was used to
normalize the CDFS observations before subtracting them from the
target observations. The cluster X-ray emission was confined to be
within $R<6$\arcmin, so that we use the annulus region outside
$R=8$\arcmin\ to probe the residual background and subtract it, taking
into account the area correction for gaps and point sources. We chose
the bin size of each annulus for the spectral extraction to be
$\ge$0\farcm5 and to have about $4000$ net counts in the 2-7keV
band. The former ensures a redistribution fraction of less than 20\%
of the flux obtained within each annulus, and the latter ensures no
more than 20\% errors in the temperature
measurements. Fig.\,\ref{f:spec} shows the X-ray spectra extracted
from the central $\le$0\farcm5 circle. The spectra were fitted with
the {\tt XSPEC} software using the combined {\tt mekal}$*${\tt wabs}
model, in which the former describes the ICM emission, and the latter
describes the Galactic absorption with a Galactic hydrogen column
density of $n_{\rm H}$. We freeze the cluster redshift to 0.2153 and
$n_{\rm H}$ to the value of $6.08\times10^{20}$\,cm$^{-2}$ given by
the LAB survey (Hartmann \& Burton 1997; Arnal et al. 2000; Bajaja et
al. 2005; Kalberla et al. 2005) at the X-ray flux-weighted centroid.

\begin{figure*}
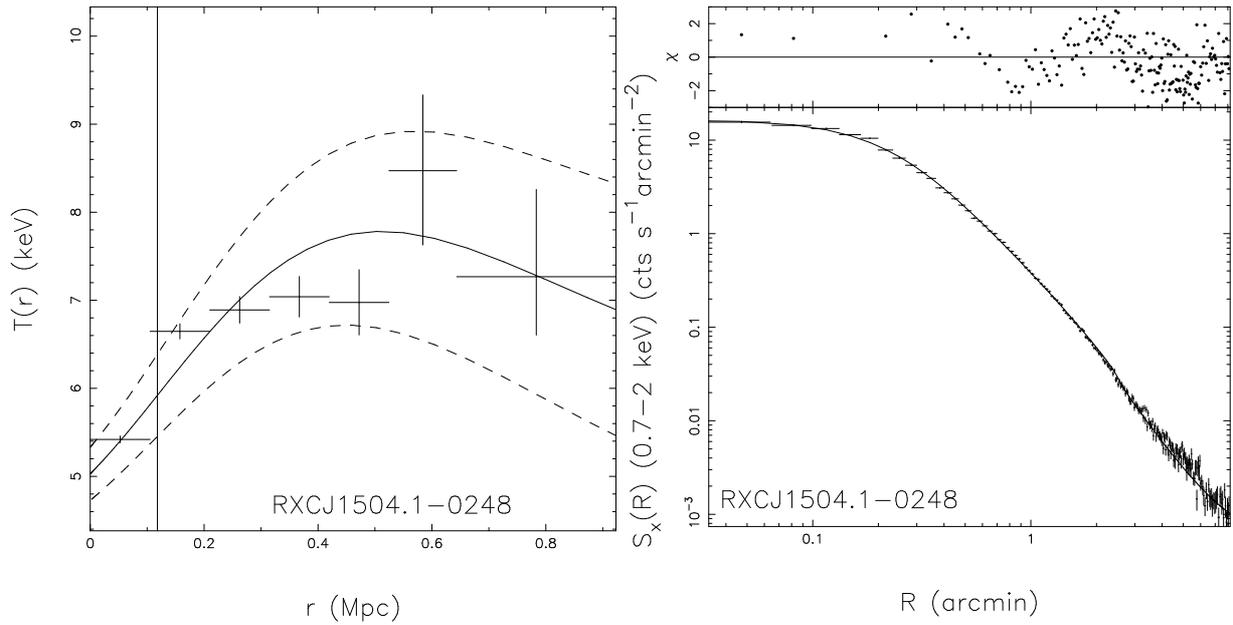

\begin{center}
\includegraphics[angle=270,width=8cm]{plots/f2a.ps}
\includegraphics[angle=270,width=8cm]{plots/f2b.ps}
\end{center}
\caption{{\it Left:} Deprojected temperature measurement versus radius
with the parametrized temperature distribution indicated by the 
solid curve and the $1\sigma$ interval by the dashed 
curves. The vertical 
line denotes $0.1r^{\rm H.E.}_{500}$. {\it Right:} Surface brightness versus 
projected distance from the X-ray flux-weighted centroid with the 
parametrized surface brightness distribution in solid curve. The 
upper panel shows deviations of the 
model from the data points normalized by the $1\sigma$ errors. 
\label{f:tneprof}}
\end{figure*}

The left panel of Fig.\,\ref{f:tneprof} shows that the radial
temperature distribution follows the universal profile of relaxed
clusters (e.g. Markevitch et al. 1998; Vikhlinin et al. 2006; Zhang et
al. 2006; Pratt et al. 2007). The observed strong cool core is also
shown in \emph{Chandra} data by B\"ohringer et al. (2005). The
amplitude of the temperature distribution is, however, 20\% lower than
that of the \emph{Chandra} measurements. Snowden et al. (2008) found
that temperature measurements with \emph{Chandra} are overestimated
for those that are hotter than 5\,keV owning to a calibration problem,
which was corrected only recently with \emph{Chandra} {\tt
  CALDB}\,v3.5.2. The temperature overestimation made by analyses of
previous \emph{Chandra} {\tt CALDB} thus accounts for the amplitude
difference between the temperature distributions measured with
\emph{Chandra} by B\"ohringer et al. (2005) and with \emph{XMM-Newton}
in this work. The right panel of Fig.\,\ref{f:tneprof} shows the
surface brightness distribution. We carried out the PSF deconvolution
and deprojection as detailed in Sect.\,2.3 of Zhang et al. (2007), in
which the \emph{XMM-Newton} PSF matrices were created according to
Ghizzardi (2001).

\subsection{Mass distribution using the hydrostatic method}
\label{s:masshe}

We assumed that the ICM is in hydrostatic equilibrium within the
gravitational potential dominated by dark matter, and that the matter
distribution is spherically symmetric. The mass could thus be
calculated from the X-ray measured radial density and temperature
distributions of the ICM by
\begin{equation} 
 \frac{1}{\mu m_{\rm p} n_{\rm e}(r)}\frac{d[n_{\rm e}(r) k
 T(r)]}{dr}= -\frac{GM(\le r)}{r^2}\,,
\end{equation} 
in which $\mu=0.62$ is the mean molecular weight per hydrogen atom,
and $k$ is the Boltzmann constant. The mass estimate was computed from
a set of input parameters $\beta$, $n_{{\rm e0}i}$, and $r_{{\rm c}i}$
($i=1,2$) representing the double-$\beta$ electron number-density
profile $n_{\rm e}(r)=n_{\rm e01}(1+r^2 r_{\rm
  c1}^{-2})^{-1.5\beta}+n_{\rm e02}(1+r^2 r_{\rm
  c2}^{-2})^{-1.5\beta}$, and $P_{i}$ ($i=1,...,7$) representing the
deprojected temperature profile, $T(r)=P_3\exp[-(r-P_1)^2
P_2^{-1}]+P_6(1+r^2 P_4^{-2})^{-P_5}+P_7$. We propagated the
uncertainties in the electron number-density and temperature
measurements using Monte Carlo simulations to compute the mass
uncertainty. The cluster radius $r_{\Delta}$ (e.g. $r_{500}$) was
defined as the radius within which the mass density is $\Delta$
(e.g. 500) times of the critical density at the cluster redshift,
$\rho_{\rm c}(z)=3H_{0}^2(8\pi G)^{-1}E^2(z)$, in which
$E^2(z)=\Omega_{\rm m}(1+z)^3+\Omega_{\Lambda}+(1-\Omega_{\rm
  m}-\Omega_{\Lambda})(1+z)^2$.

\begin{figure*}
\begin{center}
\includegraphics[angle=270,width=16cm]{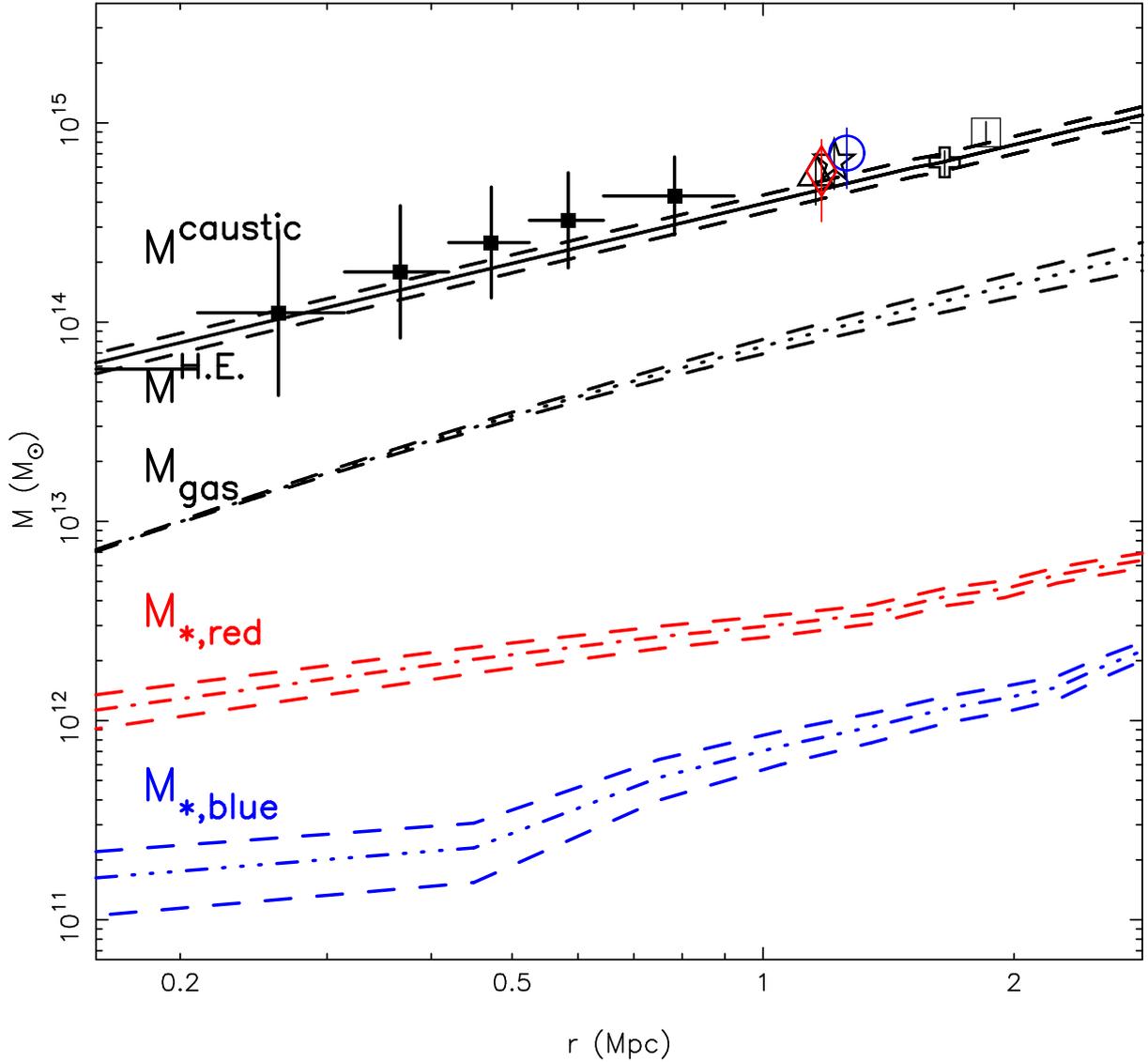}
\end{center}
\caption{Cumulative mass distributions (1) using the X-ray 
  hydrostatic method (filled squares), (2) using the caustic 
  method (black solid curve), (3) of the gas mass 
  estimate (black dotted
  curve), (4) of the stellar mass estimate of the red 
  galaxies (red dash-dotted curve),
  and (5) of the stellar mass estimate of the blue galaxies (blue
  dash-dotted-dotted-dotted curve). The dashed curves and error bars 
  denote $1\sigma$
  intervals. The open star, triangle, diamond, and circle
  indicate the dynamical mass estimates ($M_{500}^{\rm B06}$) using 53,
  48, 24 red, and 29 blue cluster members according to the Biviano et
  al. (2006) method.  The open square and cross denote the
  richness-derived mass estimates
  ($M_{200}^{\rm J07}$ and $M_{200}^{\rm H09}$)
  based on photometric data using the definitions of Johnston et
  al. (2007) and Hansen et al. (2009).
  \label{f:mprof}}
\end{figure*}

The cumulative mass distribution using the hydrostatic method is shown
in Fig.\,\ref{f:mprof} as filled squares. The cluster radius is
$r^{\rm H.E.}_{500}=(1.18\pm0.03)$\,Mpc, and the total mass estimate
is $M^{\rm H.E.}_{500}=(5.81\pm 0.49)\times 10^{14}M_{\odot}$ as
listed in Table\,\ref{t:log}. The mass profile can be well-fitted by
an NFW model (Navarro et al. 1997) with a concentration parameter of
$c_{500}=2.93\pm 0.14$.

\subsection{Two-dimensional spectrally measured maps}

We adopted the adaptive-binning methods developed by Sanders (2006)
and Cappellari \& Copin (2003; see also Diehl \& Statler 2006),
respectively, to define the masks for the 2-D analysis. We followed
the steps in Sect.\,5 of Zhang et al. (2009), except for two changes:
(i) We adopted a threshold of signal-to-noise ratio (S/N) of 30
instead of 45 to ensure an adequate number of bins in the mask. (ii)
We used the double-background subtraction strategy (i.e.  DBS II)
since there is a sufficient outer area within the \emph{XMM-Newton}
field of view (FoV) to model the residual background.

\begin{figure*}
\begin{center}
\includegraphics[angle=0,width=8.5cm]{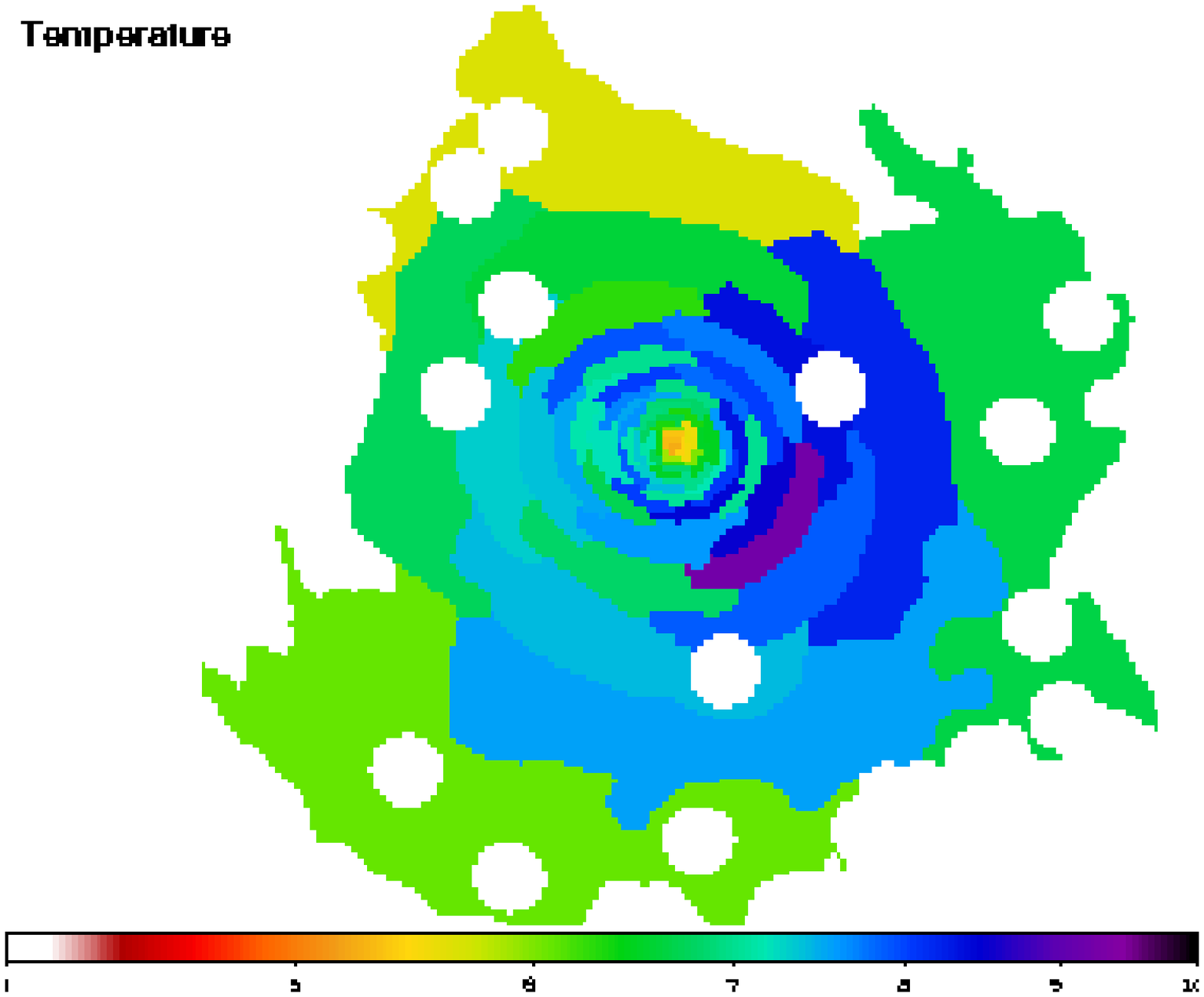}
\includegraphics[angle=0,width=8.5cm]{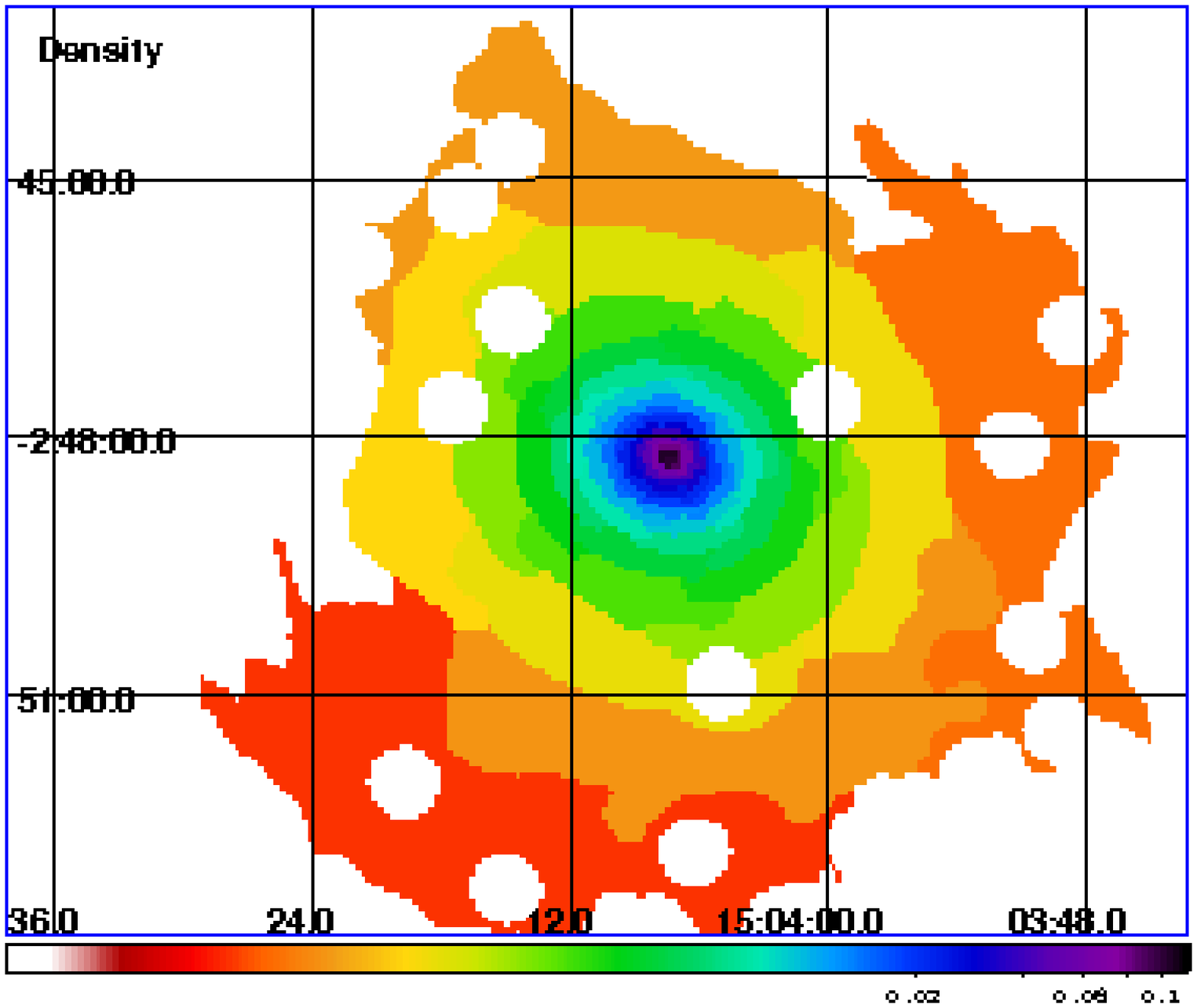}

\includegraphics[angle=0,width=8.5cm]{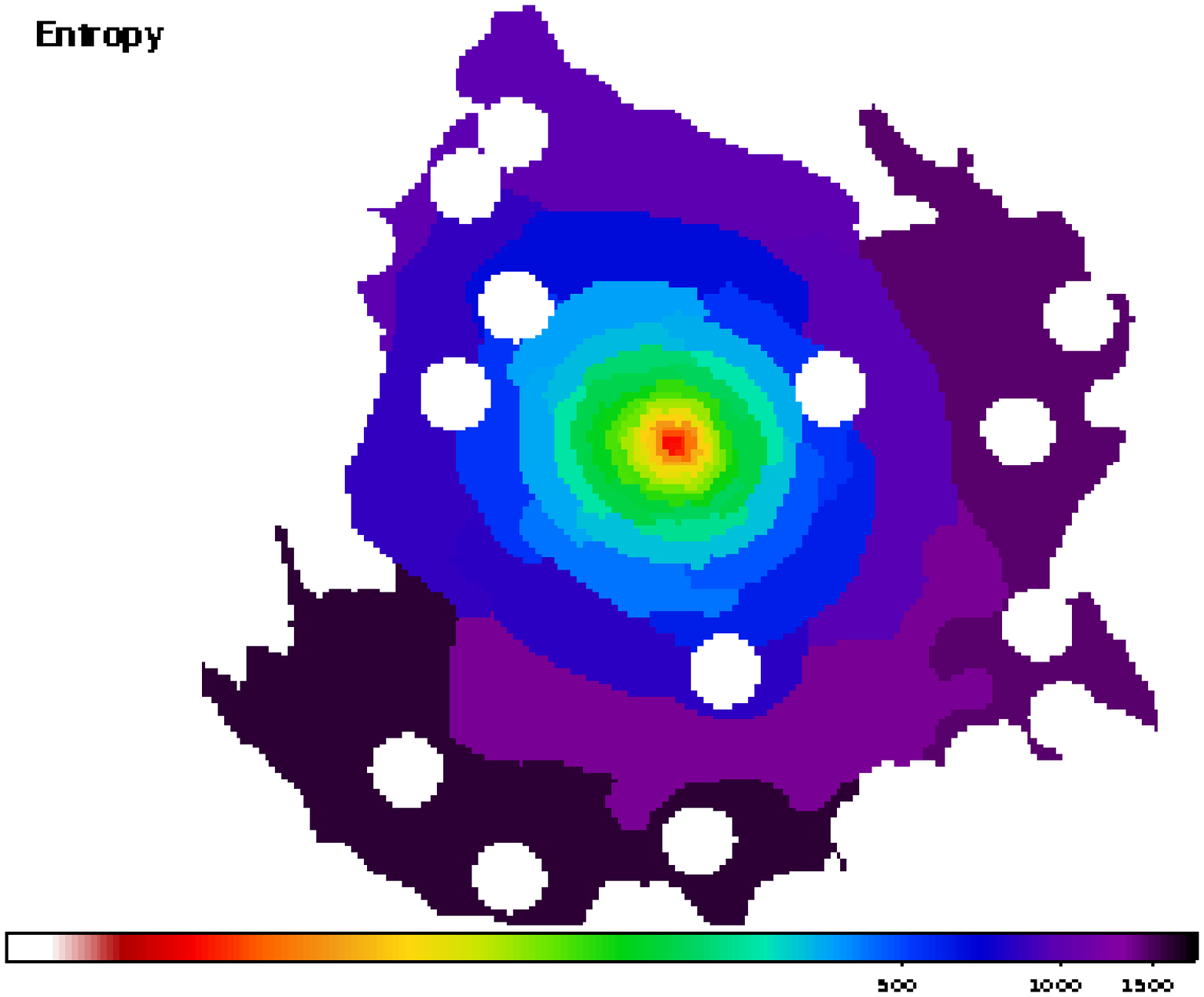}
\includegraphics[angle=0,width=8.5cm]{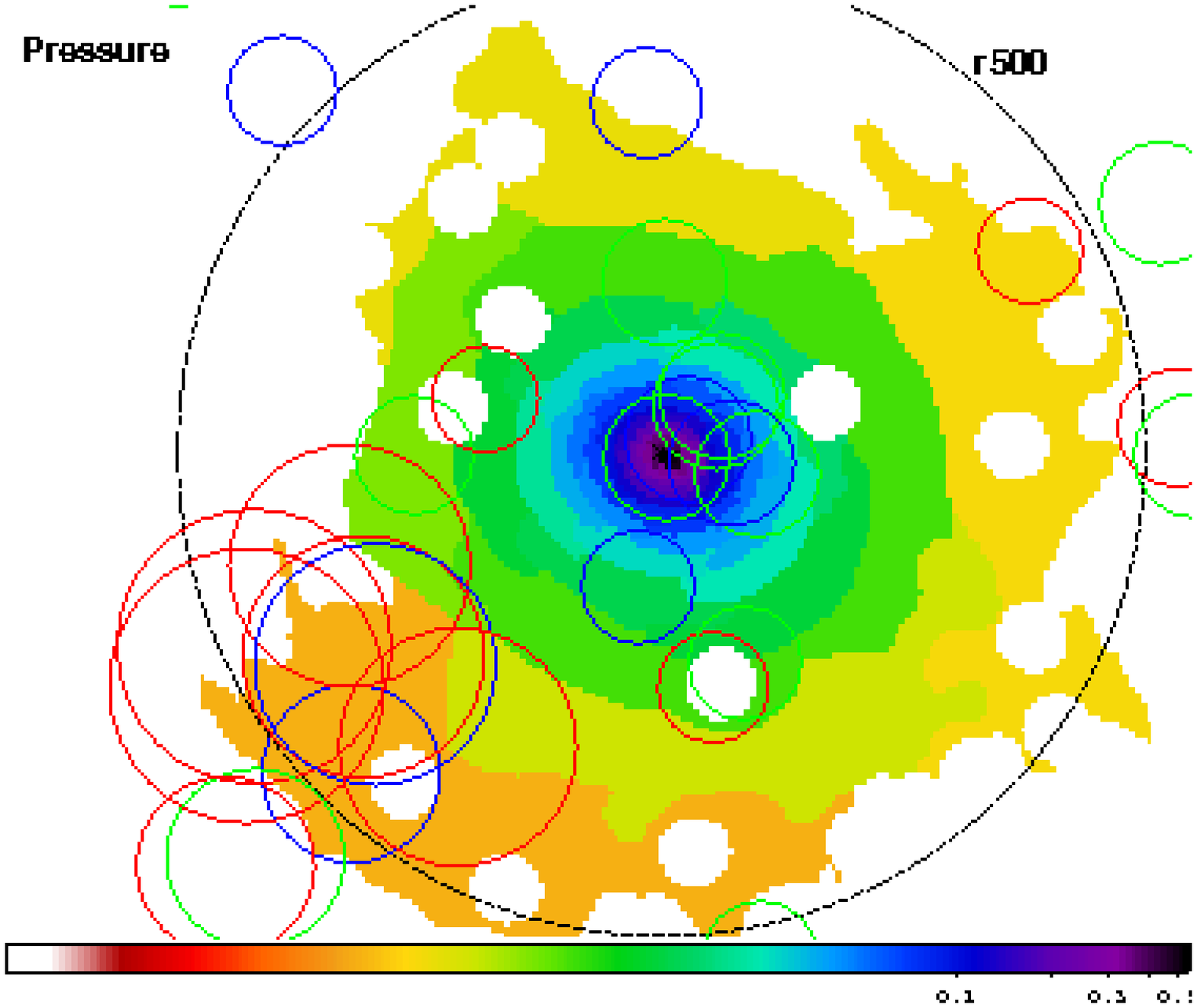}
\end{center}
\caption{Two-dimensional X-ray spectrally measured temperature (upper
  left), electron number-density (upper right), entropy (lower left),
  and pressure (lower right) maps using the Sanders (2006) binning
  scheme. Overlaid circles in blue, red, and green denote cluster
  galaxies that have spectroscopic follow-up data with their
  clustercentric velocities toward the observer greater than
  1000\,km\,s$^{-1}$, away from the observer greater than
  1000\,km\,s$^{-1}$, and smaller than 1000\,km\,s$^{-1}$,
  respectively. The radii of these circles are proportional to $\exp
  [\rho_{\rm DS}^2]$. The black cross denotes the X-ray flux-weighted
  centroid and the black dashed circle denotes $r^{\rm
    H.E.}_{500}$. The white holes mask X-ray detected point
  sources. The colour scales logarithmically from 4 to 10 in keV for
  temperature, from 0.00005 to 0.12 in cm$^{-3}$ for electron number
  density, from 10 to 1800 in keV\,cm$^2$ for entropy, and from 0.0001
  to 0.54 in keV\,cm$^{-3}$ for pressure. \label{f:2ds}}
\end{figure*}

The X-ray spectrally measured 2-D maps using the Sanders (2006) mask
are shown in Fig.\,\ref{f:2ds}. Those using the Cappellari \& Copin
(2003) mask appear in Fig.~\ref{f:2dv}. At a projected distance of
$\sim r^{\rm H.E.}_{500}$ south-east of the centroid, the surface
brightness shows a clump with its value 2.25 times the azimuthal
average of $(2.4\pm 0.4)\times 10^{-3}$~counts~s$^{-1}$~arcmin$^{-2}$,
at the same projected distance. This enables us to carry out the 2-D
spectral analysis toward larger radii in the south-east direction than
the other directions. In the temperature map, we also observe a hot
stripe at $\sim 0.3r^{\rm H.E.}_{500}$ west of the centroid.

\begin{figure*}
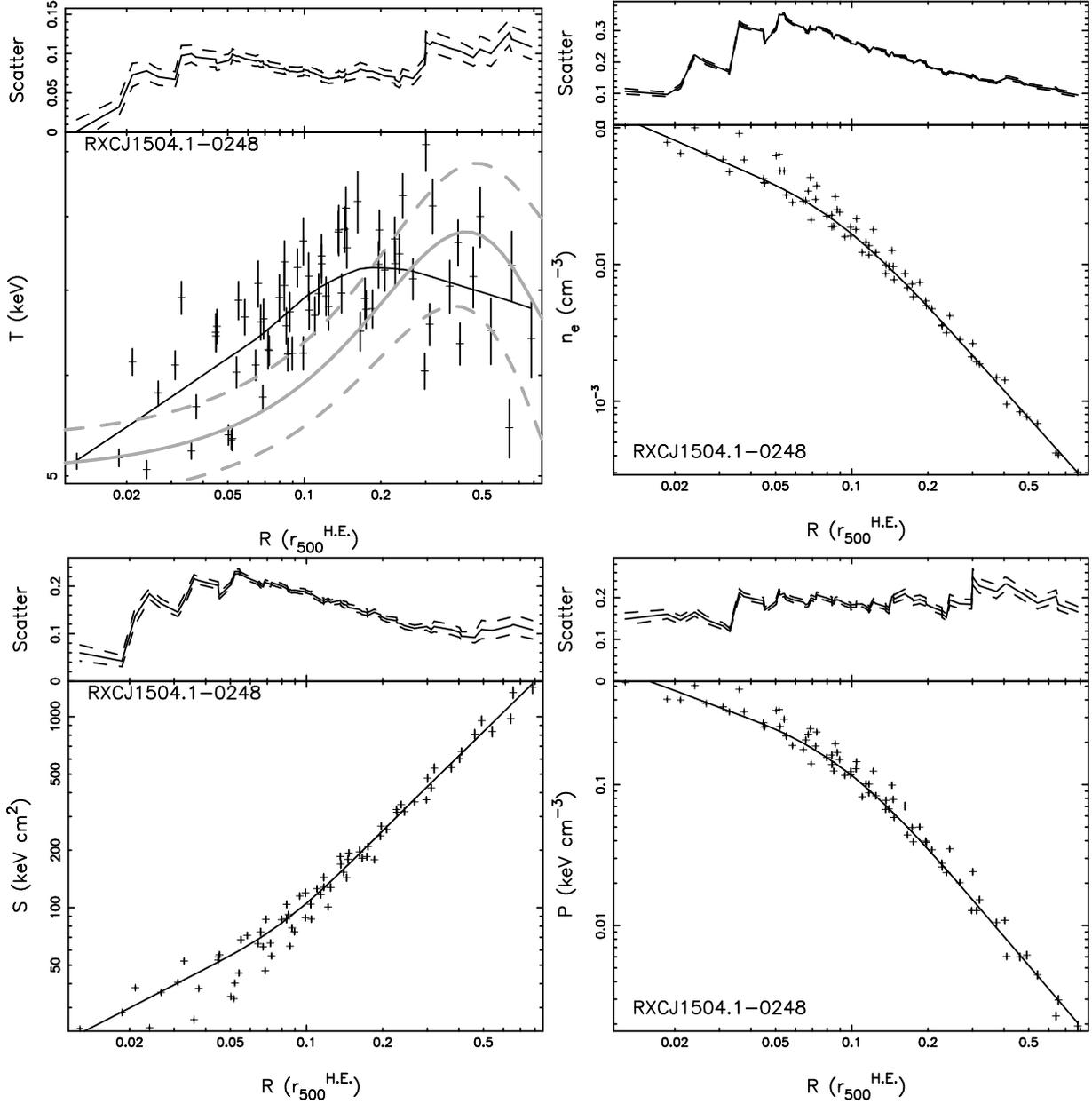

\begin{center}
\includegraphics[angle=270,width=8cm]{plots/f5a.ps}
\includegraphics[angle=270,width=8cm]{plots/f5b.ps}

\includegraphics[angle=270,width=8cm]{plots/f5c.ps}
\includegraphics[angle=270,width=8cm]{plots/f5d.ps}
\end{center}
\caption{Temperature (upper left), electron number-density 
(upper right), entropy (lower left), and pressure (lower right) 
distributions of the spectrally measured 2-D maps using the 
Sanders (2006) binning scheme are shown as crosses with 
their local 
regression fits in black curves. The solid and dashed 
grey curves in the upper left plot show the parametrization 
of the deprojected temperature profile 
and its 1-$\sigma$ interval derived in 
Sect.~\ref{s:masshe}. Differential scatter 
and its 1-$\sigma$ intervals of the temperature, electron 
number-density, entropy, and 
pressure fluctuations in the 2-D maps are shown in solid and dashed 
black curves in their upper panels. 
\label{f:scattermaps}}
\end{figure*}

We followed Zhang et al. (2009) to calculate the scatter in the
fluctuations in the 2-D maps and analyse the ICM substructure. In
Fig.~\ref{f:scattermaps}, the temperature, electron number-density,
entropy, and pressure distributions in the 2-D maps are shown as a
function of projected distance from the X-ray flux-weighted
centroid. Their non-parametric locally weighted regression
(e.g. Becker et al. 1988) was used as the mean profile. The
differential scatter and error were calculated from the area-weighted
absolute fluctuations with all bins at that projected distance. The
temperature, electron number-density, entropy and pressure maps have
scatters of smaller than $\sim$15\%, 30\%, 20\%, and 25\%,
respectively. The scatter in the temperature measurements increases
with radius, and is $\sim12$\% at the radii beyond $0.3r^{\rm
  H.E.}_{500}$. In contrast, the electron number-density measurements
appear to have a smaller scatter at large radii, which may reflect
the different relaxation timescales of the ICM temperature and
electron number density.

The 2-D maps using the Cappellari \& Copin (2003) mask demonstrate a
mild elongation along the east-west major axis on the scale of $\sim
r^{\rm H.E.}_{500}$.

\section{Optical spectroscopic data analysis  and results}
\label{s:optspe}

\subsection{VLT/VIMOS observations and data preparation}

Multi-object spectroscopy in the R1504 field was performed with the
VLT/VIMOS spectrograph using the low-resolution blue grism (LR-blue)
with the OS-blue filter, which samples the
[3700--6700]\,$\mathring{A}$ wavelength range with
$\lam/\Delta\lam=180$ (PID: 077.A-0058). The observations were carried
out in service mode, which include a bias, dark, and flat calibration
sequence with the last of these using the same set-up as the
observations.

The cluster R1504 was observed with air masses of between 1.124 and
1.172. The seeing was less than 1\farcs5 for two pointed observations
and $>2$\arcsec\ for one pointed observation. Individual targets for
the spectroscopic follow-up were selected based on their $I$-band
magnitudes in the pre-imaging data. In total, 784 slits were placed in
the 12 masks spread over three pointed observations.

Data were reduced with the {\tt VIPGI} pipeline (Scodeggio et
al. 2005). Redshifts were computed by comparing the observed spectra
to galaxy and stellar templates with the {\tt EZ} software (Garilli et
al. 2010) and improved with customized tools from Verdugo et
al. (2008) by fitting a Gaussian profile to each of the observed
strong spectral features. Not all lines, namely [O\,{\sc
  ii}]~$\lam\lam$3726,3729\,\AA\AA, [Ca\,{\sc ii}~K]~$\lam$3934\,\AA,
[Ca\,{\sc ii}~H]~$\lam$3968\,\AA, \hde~$\lam$4102\,\AA,
\hbe~$\lam$4861\,\AA, G-band ($\sim\lam$4304\,\AA), [O\,{\sc
  iii}]~$\lam\lam$4959,5007\,\AA\AA, \mgb\ ($\sim 5050-5430$\,\AA),
and [Fe\,{\sc VI}]~$\lam$5335\,\AA, are present in each spectrum. The
redshift determined from each spectrum is the mean of the individual
shifts for all visible line features. The error is the standard
deviation in these shifts, which takes into account systematic errors
in the wavelength calibration and differences in the S/N. The average
error in the individual redshift estimates is $\Delta z=0.00076$.

The R1504 field overlaps with the stellar stream of the globular
cluster Palomar-5 (e.g. Odenkirchen et al. 2001). There are thus a
large number of white- and red-dwarf stars contaminating the
spectroscopic sample. Interlopers were removed by applying the member
selection procedure in Sect.~\ref{s:specmem}.

\subsection{Spectroscopic member selection}
\label{s:specmem}

In hierarchical structure-formation scenarios, the spherical infalling
model predicts a trumpet-shaped region in the diagram of line-of-sight
velocity versus projected distance, the so-called caustic (e.g.
Kaiser 1987). The boundary defines galaxies inside the caustic as
cluster members and those outside the caustic as fore- and background
galaxies.

Several methods can select cluster members according to their
line-of-sight velocities. A constant gap or gaps weighted by projected
distances can separate member galaxies from fore- and background
galaxies (e.g. Beers et al. 1990; Girardi et al. 1993; den Hartog \&
Katgert 1996; Katgert et al. 1996, 2004; Popesso et al. 2005a; Biviano
et al. 2006). Alternatively, an adaptive kernel method, sensitive to
the presence of peaks, has been widely used to select members
(e.g. Pisani 1993, 1996; Fadda et al. 1996; Diaferio \& Geller 1997;
Diaferio 1999; Geller et al. 1999; Rines \& Diaferio 2006).

We modified the adaptive kernel method of Diaferio (1999) for
medium-distance clusters (e.g. Harrison \& Noonan 1979; Danese et
al. 1980), and used it to identify the interlopers as follows.
Galaxies with redshifts $ | cz - cz_{\rm cl} | \le 4000$\,km\,s$^{-1}$
were preliminarily selected and located in the $(r,v)$ diagram, where
$r$ is the galaxy transverse separation from the cluster centroid, and
$v$ is the line-of-sight velocity of the galaxy relative to the
cluster. At the redshift of R1504, we modified the calculation of
$(r,v)$ to $r=D_{\rm a}\sin\theta$ and $v=(cz-cz_{\rm
  cl}\cos\theta)/(1+z_{\rm cl})$, where $D_{\rm a}$, $\theta$, $z$,
and $z_{\rm cl}$ are the angular diameter distance of the cluster,
angular separation of the galaxy from the cluster centroid, galaxy
redshift, and cluster redshift. The angular diameter distance is
defined as $D_{\rm a}=c H_{0}^{-1} (1+z_{\rm cl})^{-1}
\int_{0}^{z_{\rm cl}} E^{-1}(z)dz$. We scaled $r$ and $v$ to ensure
approximately equal weights in $r$ and $v$, and computed the 2-D
adaptive density distribution $f(r,v)$. The boundary of the caustic is
defined by $f(r,v)=\kappa$, in which $\kappa$ is determined by
minimizing $\vert \langle v_{\rm esc}^2\rangle_{\kappa,R} - 4\langle
v^2\rangle_R\vert^2$. Here $\langle v_{\rm esc}^2\rangle_{\kappa,R}=
\int_0^R{\cal A}^2(r)\varphi(r)dr/\int_0^R\varphi(r)dr$ is the escape
velocity, $R$ and $\langle v^2\rangle_R$ are the mean projected
distance and velocity dispersion of the members, and $\varphi(r)=\int
f(r,v)dv$. In addition, ${\cal A}(r)=\min\{\vert v_{\rm u}\vert,\vert
v_{\rm d}\vert\}$, which represents the minimum of the upper and lower
solutions of $f(r,v)=\kappa$.

Fig.\,\ref{f:caustic} shows the histogram of the line-of-sight
velocity and the diagram of line-of-sight velocity versus projected
distance of galaxies. The 53 galaxies inside the caustic boundary are
considered as members. Our further dynamical analysis is based on
these 53 spectroscopic member galaxies.

\subsection{Mass distribution using the caustic method}
\label{s:caustic}

We applied the biweight estimator developed by Beers et al. (1990) to
measure the redshift and velocity dispersion, in which the errors are
estimated by means of 1000 bootstrap simulations. The cluster redshift
was assumed to be the biweight estimator of location of the
recessional velocities, $cz_{i}$. The velocities in the cluster
  reference frame $v_{{\rm rest},i}$ were derived from the recessional
  velocities $cz_{i}$ using the formula $v_{{\rm
      rest},i}=(cz_{i}-cz_{\rm cl})(1+z_{\rm cl})^{-1}$. The velocity
dispersion was given by the biweight estimator of scale of these
$v_{{\rm rest},i}$ values. With all 53 spectroscopic members, the
measured redshift is ($0.2165\pm 0.0005$), and the velocity dispersion
is ($1132\pm 94$)\,km\,s$^{-1}$.

The escape velocities of galaxies extracted from the amplitude of the
caustic scale with the gravitational potential of the dark matter halo
(e.g. Diaferio \& Geller 1997; Diaferio 1999; Rines \& Diaferio 2006)
\begin{equation}
GM(\le r) = {1\over 2} \int_0^r {\cal A}^2(x) dx.
\label{e:caustic}
\end{equation}
The error in the cumulative mass estimate within $r_i$ is defined as
$\delta M_{i}=\sum_{j=1}^{i}|2m_j \delta{\cal A}(r_j)/{\cal A}(r_j)|$,
in which $m_j$ is the mass estimate in the $j^{\rm th}$ shell,
$\delta{\cal A}(r)/{\cal A}(r)=\kappa/\max\{f(r,v)\}$, and
$\max\{f(r,v)\}$ is the maximum value of $f(r,v)$ at fixed $r$. Since
the caustic of a spherical system is symmetric, the minimum of the
upper and lower caustics is taken as the caustic amplitude at each
radius, which excludes interlopers more effectively than taking the
average of the upper and lower caustics as the caustic amplitude. The
caustic method does not assume dynamical equilibrium, and is thus able
to measure cluster masses out to large radii. The mass distribution
using the caustic method is shown in Fig.~\ref{f:mprof}, and implies
that $r^{\rm caustic}_{500}=(1.06\pm 0.04)$~Mpc (Table\,\ref{t:log}).

\subsection{Two-dimensional kinematic structure}

As clusters of galaxies formed at moderate look-back times, they often
contain substructures caused by the accretion of and merging with
smaller systems. The quantification of substructures is thus important
to assess the reliability of different mass measurements and infer the
dynamical history of the cluster. It is, however, difficult to
quantify substructures at large radii with \emph{XMM-Newton} data
because of the high background level relative to the faint X-ray
emission near $r^{\rm H.E.}_{500}$. Optical spectroscopic data provide
a promising alternative means of quantifying substructures at all
radii.

Dressler \& Shectman (1988, DS hereafter) developed a statistical
method for detecting substructures in galaxy clusters related to
systematic deviations from the average spatial and/or velocity
structure, which we summarize as follows. The deviation in the local
velocity mean and dispersion from the global mean and dispersion of
the cluster is defined as
\begin{equation}
\rho_{\rm DS}=\sqrt{\frac{N_{\rm local}+1}{ \sigma^{2}} [(\bar{v}_{\rm
  local}-\bar{v})^2+(\sigma_{\rm local}-\sigma)^2]}\,.
\end{equation}
The local line-of-sight velocity and dispersion, $\bar{v}_{\rm local}$
and $\sigma_{\rm local}$, are measured with its $N_{\rm local}=10$
nearest neighbours in projection. The global line-of-sight velocity
and dispersion, $\bar{v}=64950$\,km\,s$^{-1}$ and
$\sigma=1132$\,km\,s$^{-1}$, are given in Sect.~\ref{s:caustic} as
measured for all (i.e. $N_{\rm all}=53$ for R1504) spectroscopic
cluster members. A cumulative deviation $\Delta_{\rm
  DS}=\sum_{i=1}^{N_{\rm all}}\rho_{{\rm DS},i}$ is of the
order of $N_{\rm all}$ when the line-of-sight velocity obeys a
Gaussian distribution and local variations are only random
fluctuations, but differs from $N_{\rm all}$ when the line-of-sight
velocity distribution deviates from a Gaussian. Therefore, the
$\Delta_{\rm DS}$ statistic is sensitive to the presence of
substructures. A higher $\Delta_{\rm DS}$ value indicates a higher
possibility of containing substructures.

In Fig.\,\ref{f:2ds}, we display the 53 spectroscopic galaxies by
circles with their radii proportional to $\exp [\rho_{\rm DS}^2]$. The
blue, red, and green colours denote cluster galaxies that have
spectroscopic follow-up data with their clustercentric velocities
toward the observer greater than 1000\,km\,s$^{-1}$, away from the
observer greater than 1000\,km\,s$^{-1}$, and smaller than
1000\,km\,s$^{-1}$. The DS test yields $\Delta_{\rm DS,obs}=86.8$, and
reveals a clustering of galaxies with large $\rho_{\rm DS}$ values at
a projected distance of $\sim r^{\rm H.E.}_{500}$ south-east of the
cluster centroid. This may be indicative of a relatively high-velocity
($>1000$~km\,s$^{-1}$) group associated with the main cluster.

To estimate the robustness of the $\Delta_{\rm DS}$ statistic, we
carried out $n_{\rm MC}=10,000$ Monte Carlo simulations by randomly
shuffling the line-of-sight velocities of the member galaxies.  This
procedure erases any true correlations between the line-of-sight
velocities and positions. The probability of having substructures can
be estimated by $P_{\rm DS}= \sum_{j=1}^{n_{\rm MC}}\delta_{j}/n_{\rm
  MC}$, in which $\delta_{j}=\Delta_{\rm DS,j}$ when $\Delta_{\rm
  DS,j}>\Delta_{\rm DS,obs}$ and $\delta_{j}=0$ otherwise (e.g. Hou et
al. 2012). Therefore, systems with significant substructures show
low-$P_{\rm DS}$ values. The Monte Carlo realization of the
$\Delta_{\rm DS}$ statistic yields $P_{\rm DS}=0.06$ for R1504, which
only corresponds to a $\sim10$\% false detection according to Hou et
al. (2012). This $P_{\rm DS}$-value thus represents a non-negligible
possibility that the high velocity group exists. The histogram of the
line-of-sight velocities of possible member galaxies of the high
velocity group is shown as (red) dashed lines in the left panel of
Fig.\,\ref{f:caustic}.

\section{Optical imaging data analysis and results}
\label{s:optima}

\subsection{WFI observations and data preparation}

The cluster R1504 was observed with the ESO/MPG 2.2\,m telescope with
the WFI (Baade et al. 1999) in the $B/123$-, $V/89$-, and
$R_{c}/162$-bands (hereafter $B$-, $V$-, and $R$-bands) during three
runs in June 2009, 2010, and 2011 mostly under non-photometric
conditions. The WFI has a $4 \times 2$ mosaic detector of ${\rm 2k
  \times 4k}$ CCDs, which has a pixel size of 0\farcs238 and a FoV of
34\arcmin$\times$33\arcmin. We reduced the data and co-added the
single exposures using the {\tt THELI} pipeline (Erben et
al. 2005). The coadded $B$-, $V$-, and $R$-band images have total
exposure times of 9238\,s, 8317\,s, and 11035\,s with 1\farcs27,
1\farcs21, and 0\farcs96 seeing. Fig.\,\ref{f:wfi} shows the WFI
$B$-band image.

Since the observing conditions of the imaging varied slightly,
photometric calibration of the WFI $R$-band data was tied to the
shallower SDSS $r$-band data in the same field. The $B$- and $V$-band
images were calibrated against the $R$-band by fitting the observed
main sequence of stars with that of a photometrically calibrated
field. Cross-checks were applied by comparing predicted with observed
colours for galaxies with spectroscopic redshifts using procedures
typical of photometric redshift codes. The absolute calibration of the
photometric zero point is more accurate than 0.1~magnitude. The
internal colour calibration is more accurate than 0.04~magnitude.

Photometry was carried out with {\tt SExtractor} (Bertin \& Arnouts
1996) in dual mode with the $R$-band image set as the detection
frame. Galaxy colours were measured in an aperture of 2\farcs5
diameter.

\subsection{Photometric member selection}

The broad $B$- and $R$-bands can probe the spectral range across the
4000\,$\mathring{A}$ break for galaxy clusters at $z\sim 0.2-0.3$,
which can be used to characterize the bulk of stellar
populations. Early-type galaxies display red colours and reside in a
narrow band, the so-called red sequence (e.g. Baum 1959), in the
colour--magnitude diagram. The $B-R$ colour versus $R$-band magnitude
is shown in the left panel of Fig.\,\ref{f:rslf} for all galaxies
within $r^{\rm H.E.}_{500}$, in which we highlight the 29 blue and 24
red spectroscopic members regardless of their clustercentric
distances.

We fit the red sequence with a robust biweight linear-fitting
algorithm, which is negligibly affected by skewness (e.g. Beers et
al. 1990; Gladders et al. 1998). The best-fit relation is $B-R =
-0.097R + 3.88$ with a scatter of $\sigma=0.083$\,magnitude.

We define red members as galaxies within $3\sigma$ of the best-fit red
sequence. Galaxies that are more than $3\sigma$ bluer than the red
sequence are considered as blue galaxies, whereas galaxies that are
more than $3\sigma$ redder than the red sequence are discarded because
they are most likely background galaxies. We also discard galaxies
bluer by more than 1.5 magnitude than the red sequence to reduce
foreground contamination. These limits are marked in
Fig.\,\ref{f:rslf}.

The comparison between the galaxy number counts in the R1504 field and
the Garching-Bonn Deep Survey (GaBoDS, Erben et al. 2005; Hildebrandt
et al. 2006) fields shows that our photometric sample is 100\%
complete down to $R = 22$\,magnitude. This limit corresponds to
$M_{\rm R}\approx -18$\,magnitude for galaxies of different colours at
$z_{\rm cl}=0.2153$, and is indicated by an oblique curve in
Fig.\,\ref{f:rslf}. The {\tt CLASS\_STAR} parameter provided by {\tt
  SExtractor} is completely reliable down to $R=21.5$\,magnitude and
90\% reliable down to $R=22$\,magnitude. We thus limit the analysis to
the galaxies brighter than $R = 22$\,magnitude. The limiting magnitude
is $R=24.44$\,magnitude at $5\sigma$, which was calculated in an
aperture of 1\arcsec\ diameter. The stellar contamination should be
quite low in the photometric sample because of the high quality of the
WFI imaging.

\subsection{Galaxy luminosity function and stellar mass}

To determine the photometric properties of the cluster population, we
corrected for contamination by both fore- and background galaxies
along the line-of-sight by applying a statistical subtraction using a
subset (2.25\,deg$^2$) of the GaBoDS, whose photometric data are
complete in all bands. We applied the same magnitude and colour cuts
to the GaBoDS fields as we did to the R1504 field, and constructed the
galaxy number counts as a function of magnitude for red and blue
galaxies, respectively. The galaxy number counts in the GaBoDS fields
were normalized to the R1504 sky area and subtracted from the
corresponding galaxy number counts in the R1504 field. The size of the
magnitude bin used to construct the galaxy number counts was chosen to
be large enough to ensure that there were positive counts in all
magnitude bins after subtracting the background. The errors are a
combination of the Poisson statistical error and cosmic variance. The
latter was measured to be the standard deviation in the galaxy number
counts across the GaBoDS fields. The Poisson statistical error
dominates the error budget at inner radii because there are relatively
fewer objects, whereas cosmic variance dominates at large radii.

The right panel of Fig.\,\ref{f:rslf} shows the $R$-band galaxy
luminosity functions for red (red), blue (blue), and all (black)
members within $r^{\rm H.E.}_{500}$. We derived the best-fit Schechter
(1976) functions for red and all members, respectively, with a
$\chi^2$ minimization algorithm. The red members show a shallower
slope at the faint end than that of all members. As listed in
Table\,\ref{t:log}, both the characteristic $R$-band magnitude,
$R^{\ast}$, and the slope at the faint end, $\alpha$, agree with that
of nearby clusters (e.g. Christlein \& Zabludoff 2003; Popesso et
al. 2005b; Durret et al. 2011). No fit was possible for blue galaxies.

The brightest cluster galaxy (BCG) is very blue ($B-R=0.98$) and
bright ($R=16$~magnitude). Since the BCG cannot normally be described
by the Schechter function, we explicitly removed the BCG from the
photometric sample in the above calculation of the galaxy luminosity
function. The total $R$-band luminosity of the cluster was computed by
explicitly adding the BCG luminosity to the integration of the galaxy
luminosity function. We tested how the BCG colour affects the measured
optical luminosity by adding the BCG luminosity to the red and blue
populations, respectively. The resulting uncertainty in the $R$-band
luminosity is smaller than 3\%. The measured $R$-band optical
luminosity within $r^{\rm H.E.}_{500}$ is listed in
Table\,\ref{t:log}.

We adopted the stellar mass-to-optical light ratio from Table\,7 of
Bell et al. (2003), $\log(M_{\ast}/L_{R}) = a + b(B-R)$, in which $a=
-0.523$, $b= 0.683$, $B-R$ is the rest-frame colour, and a ``diet''
Salpeter (1955) stellar initial mass function (Bell \& de Jong 2001)
is adopted. We show the cumulative stellar mass profiles for red and
blue galaxies, respectively, in Fig.\,\ref{f:mprof}. The stellar mass
estimates for blue, red, and all photometric members within $r^{\rm
  H.E.}_{500}$ are given in Table\,\ref{t:log}. The cluster R1504 has
a rather low stellar-mass fraction among nearby clusters (e.g. Biviano
\& Salucci 2006), although there is an agreement given the large
scatter for nearby clusters (e.g. Andreon 2010; Zhang et
al. 2011b). The contribution to the total stellar mass budget
associated with intra-cluster light (e.g. Zibetti et a. 2005) could
reach 10--20\% level at the mass scale of galaxy groups (e.g. Gonzalez
et al. 2007). The difference in the $R$-band luminosity caused by
adding the BCG luminosity to the blue and red populations,
respectively, is within 3\% for R1504. Therefore, both the
intra-cluster light and BCG play non-dominant roles to explain the low
stellar-mass fraction.

\subsection{Blue-galaxy fraction}

Understanding rapid galaxy-morphology transformation is essential to
model the mass assembly history of galaxy clusters. Red galaxies with
low levels of ongoing star formation are assumed to be the descendants
of blue star-forming galaxies that have been accreted from the
surrounding filamentary structure and quenched by specific processes
operating within the cluster environment (e.g. Braglia et al. 2007;
Verdugo et al. 2012). How spirals were transformed into ellipticals
remains unclear (e.g. Poggianti \& Wu 2000). The efficiency of
different transformation processes varies with environment. Ram
pressure stripping, for instance, is more effective in the cluster
core where there is dense ICM (e.g. Fujita \& Nagashima 1999; Quilis
et al. 2000). Galaxy-galaxy merging, on the other hand, is more
effective in infalling groups at the cluster outskirts (e.g. Kauffmann
et al. 1999).

In the left panel of Fig.\,\ref{f:wfiprof}, we present the number
density profiles of the red and all member galaxies, respectively, as
a function of projected distance from the X-ray flux-weighted
centroid. Similar to the findings for other clusters (e.g. Lin et
al. 2004), the red population in R1504 is well-fitted by a projected
NFW profile (Navarro, Frenk \& White 1997; Bartelmann 1996), in which
the reduced $\chi^2$ is 1.03. The fit of a projected NFW profile to
all members is relatively poor with $\chi^2 = 2.24$. The right panel
of Fig.\,\ref{f:wfiprof} displays the blue-galaxy fraction. Within a
projected distance of 700\,kpc, where the ICM is dense, the galaxy
population is dominated by red galaxies with a fraction larger than
70\%. Beyond 700\,kpc, the blue-galaxy fraction first increases
rapidly, reaching 60\% at $\sim r^{\rm H.E.}_{500}$, and then stays
almost constant in the outskirts.

The large error bars in the blue-galaxy fraction at large radii take
into account the uncertainty in the background. The colours of the
field galaxies have their own variation. One cannot exclude that the
high blue-galaxy fraction is coincidently due to a larger field-galaxy
contamination. A reliable measure of the blue-galaxy fraction free of
background contamination requires extensive
spectroscopy. Nevertheless, the high blue-galaxy fraction beyond
$r^{\rm H.E.}_{500}$ may indicate a critical distance where specific
processes in clusters such as starvation (e.g. Balogh et al. 2000)
start to affect the properties of infalling galaxies. This distance
may, however, depend on the environment (e.g. Urquhart et
al. 2010). Urquhart et al. (2010, Figure\,10 therein) pointed out an
excess of extremely blue galaxies in massive groups, in which active
starburst galaxies are driven by galaxy-galaxy interactions in the
group environment. The radial blue-galaxy fraction and presence of the
high velocity group in R1504 support this interpretation.

\subsection{Richness-derived mass estimates}
\label{s:mphoto}

There are a number of approaches for calculating the cluster total
mass based on photometric measurements (e.g. Hansen et al. 2005, 2009;
Johnston et al. 2007; Reyes et al. 2008).

We can iteratively estimate the mass from the richness distribution
according to existing scaling calibrations between richness and mass
estimates at a fixed overdensity. The cluster total mass is measured
within the radius where the mass density is 200 times the critical
density in Johnston et al. (2007) and Hansen et al. (2009), but 200
times the mean density of the Universe in Hansen et al. (2005) and
Reyes et al. (2008). To be consistent with the mass definition, we use
the calibrations between richness and weak-lensing mass measurements
in Johnston et al. (2007), eq.~(26), and Hansen et al. (2009),
eq.~(10), respectively, to calculate the richness-derived mass
estimates. In these calibrations, the richness was measured down to
$0.4L^{\ast}$ in the SDSS $i$-band, which is equivalent to
$(i^{\ast}+1)$ magnitude. Since there were no $i$-band data for R1504,
we used the $R$-band derived richness down to $(R^{\ast}+1)$ to
approximate their $i$-band derived richness down to
$(i^{\ast}+1)$. Nevertheless, the characteristic magnitude could be
taken as a characteristic length-scale for galaxy formation at a
similar cosmic epoch. Therefore, galaxies brighter than $(i^{\ast}+1)$
overlap significantly with those brighter than $(R^{\ast}+1)$. The
richness-derived mass estimates (see Table\,\ref{t:log}) agree well
with the hydrostatic and dynamical mass measurements.

\section{Discussion}
\label{s:mcom}

\subsection{Uncertainties in the dynamical mass estimates}

To quantify the uncertainties in the dynamical mass estimate due to
substructures and infalling blue galaxies, we need a method to derive
the dynamical mass estimate based on a small number of member
galaxies. A different method from the caustic one, which was detailed
in Sect.~3 of Biviano et al. (2006) based on a sample of simulated
clusters, serves this purpose, and is summarized as follows.

We measured the velocity dispersion ($\sigma_{\rm a,p}$ in Biviano et
al. 2006) with a number of member galaxies defined according to
Sect.~\ref{s:msub} and Sect.~\ref{s:blue}, respectively. An initial
estimate of the mass was derived using eq.~(2) in Biviano et
al. (2006), namely $M_{\rm v} \equiv A [{\sigma_{\rm v} / (10^3 {\rm
    \,km\,s^{-1}})}]^3\times 10^{14} h^{-1} E^{-1}(z) M_{\odot}$ with
$A=1.50 \pm 0.02$ under the assumption of $\sigma_{\rm
  v}=\sqrt{3}\sigma_{\rm a,p}$. An estimate of the corresponding
cluster radius, $\tilde{r_{\rm v}}$, was derived by following steps 7
and 9 in Biviano et al. (2006). Replacing the true quantities $r_{\rm
  v}$ and $\sigma_{\rm a}$ with their estimates $\tilde{r_{\rm v}}$
and $\sqrt{3}\sigma_{\rm a,p}$ in Fig.~4 in Biviano et al. (2006), we
obtained an improved estimate of $\sigma_{\rm v}$. The final cluster
mass estimate ($M^{\rm B06}$ hereafter) was calculated using eq.~(2)
in Biviano et al. (2006) with the improved estimate of $\sigma_{\rm
  v}$. The mass error was derived by combining in quadrature the error
in the velocity dispersion converted to mass, the additional error
introduced by the uncertainty in Fig.~4 in Biviano et al. (2006), and
the 10\% mass systematic according to the Biviano et al. (2006)
sample. We computed $M^{\rm B06}_{500}$ from $M^{\rm B06}_{200}$ with
the NFW model, in which the concentration parameter is given in step 7
of Biviano et al. (2006) as $c^{\rm B06}=4 [\sigma_{\rm a,p}/(700{\rm
  \,km\,s^{-1}})]^{-0.306}$.

\subsubsection{Uncertainty due to substructures}
\label{s:msub}

The high velocity group revealed by the DS test overlaps with the
substructure in the 2-D X-ray maps. In the high-resolution
weak-lensing shear map by Klein et al. (in prep.) based on Subaru
observations, there is a main lensing peak at the BCG position
additional to a secondary peak near the high velocity group. The
elongation of the surface mass density follows the elongation shown in
the X-ray 2-D maps.

The dynamical mass measurement overestimates the total mass of a
cluster with substructures (e.g. Biviano et al. 2006). We quantified
the uncertainty in the dynamical mass estimate due to the
high velocity group as follows. The velocity dispersion is ($1132\pm
94$)\,km\,s$^{-1}$ based on all 53 spectroscopic members. Excluding
the five galaxies belonging to the high velocity group, which are
shown as large red circles at $\sim r^{\rm H.E.}_{500}$ south-east of
the cluster centroid in Fig.\,\ref{f:2ds}, the velocity dispersion is
reduced to ($1079\pm 97$)\,km\,s$^{-1}$. As shown in
Table\,\ref{t:log} and Fig.~\ref{f:mprof}, the total mass estimate is
$M^{\rm B06}_{\rm 500,53m}=(6.355\pm 1.711)\times 10^{14} M_{\odot}$
based on the 53 members and $M^{\rm B06}_{\rm 500,48m}=(5.452\pm
1.573)\times 10^{14} M_{\odot}$ based on the 48 members excluding the
high velocity group, following the Biviano et al. (2006) method. The
mass uncertainty caused by the substructure of the high velocity group
is $\sim$14\%. Excluding the high velocity group improves the
agreement between the X-ray hydrostatic and dynamical mass
estimates. This highlights the importance of a proper analysis of the
structures of evolving systems such as cluster of galaxies in order to
obtain robust mass measurements.

\subsubsection{Mass discrepancy using red and blue galaxies}
\label{s:blue}

For a cluster in dynamical equilibrium, both blue and red galaxy
populations behave similarly in the diagram of line-of-sight velocity
versus projected distance, which reflects the same underlying mass
distribution. For a cluster with a rich infalling population, the blue
population tends to overestimate the total mass since the infalling
members bias the velocity dispersion toward high values (e.g. Carlberg
et al. 1997). Fifty-five percent of the spectroscopic member galaxies
are blue galaxies. To quantify the uncertainty in the dynamical mass
estimate due to infalling blue galaxies, we derived the velocity
dispersion and dynamical mass estimates for blue members and red
members separately. The blue population has a velocity dispersion of
($1170\pm 125$)\,km\,s$^{-1}$, which yields a cluster mass estimate of
$M^{\rm B06}_{\rm 500,blue}=(7.057\pm 2.373)\times 10^{14} M_{\odot}$
according to Biviano et al. (2006). The red population has a velocity
dispersion of ($1096\pm 156$)\,km\,s$^{-1}$, which yields a cluster
mass estimate of $M^{\rm B06}_{\rm 500,red}=(5.722\pm 2.517)\times
10^{14} M_{\odot}$. The dynamical mass estimate derived from red
members, $M^{\rm B06}_{\rm 500,red}$, is in closer agreement with the
hydrostatic mass estimate than that using all members. The cluster
mass estimate based on the blue population, $M^{\rm B06}_{\rm
  500,blue}$, is 1.23 times the mass estimate based on the red
population (Table\,\ref{t:log} and Fig.~\ref{f:mprof}). It indicates
that R1504 may not contain a rich infalling population, which is also
supported by the caustic comparison between R1504 and the simulated
clusters (e.g. Figure~6 of Gill et al. 2005).

\subsection{Hydrostatic versus dynamical versus photometric mass
  estimates}

We derived the mass distribution using the hydrostatic method based on
X-ray imaging spectroscopy and the dynamical method based on optical
spectroscopy, independently (Fig.\,\ref{f:mprof}). At $r^{\rm
  H.E.}_{500}$, the mass estimate from the caustic method is $(4.66
\pm 0.47) \times 10^{14}M_{\odot}$, which is 80\% of the X-ray
hydrostatic mass estimate (Table\,\ref{t:log}). The discrepancy
between the hydrostatic and dynamical mass estimates is comparable to
the mass discrepancy between the results obtained using red and blue
galaxies. In addition, the 2-D pressure map shows a $\sim$20\%
scatter, which can be propagated into the hydrostatic mass
estimate. Despite the mass uncertainties, the X-ray hydrostatic mass
estimate is slightly higher than the dynamical one at $r^{\rm
  H.E.}_{500}$ because the mass profile of the latter appears to be
more concentrated than the former.

As shown in Table\,\ref{t:log}, the richness-derived mass estimates
according to recent calibrations (e.g. Johnston et al. 2007; Hansen et
al. 2009) agree well with the hydrostatic and dynamical mass
estimates.

It is also important to compare the mass measurements at fixed
overdensities (e.g. 500 and 200) as these masses are used in
cosmological applications. At the overdensity of 500, the mass
estimate from the caustic method is 72\% of the hydrostatic mass
estimate. This is due to the less-concentrated mass distribution
determined using the hydrostatic method than that determined from the
dynamical method, which can be caused by the disturbed ICM in the
cluster core because of AGN feedback as discussed in
Sect.~\ref{s:core}. At the overdensity of 200, the richness-derived
mass estimates according to Hansen et al. (2009) and Johnston et
al. (2007) are about 0.8--1.2 times the hydrostatic mass estimate and
1.0--1.4 times the dynamical mass estimate.

\subsection{Agreement with the scaling relations}

The X-ray properties of R1504 are tabulated in Table\,\ref{t:log}. The
global temperature and metallicity are measured with the
\emph{XMM-Newton} spectra extracted from the $(0.2-0.5)r^{\rm
  H.E.}_{500}$ annulus. The X-ray luminosity was derived by
integrating the surface brightness out to $r^{\rm H.E.}_{500}$. We
show the X-ray luminosity of the 0.1--2.4\,keV band and 0.01--100~keV
(bol hereafter) band, respectively, in the rest frame including the
cool core (incc hereafter) and correcting for the cool core (cocc
hereafter), respectively. The X-ray luminosity corrected for the cool
core is computed by assuming a constant value of the surface
brightness distribution in the cluster core, $S^{\rm cocc}_{\rm X}(R <
0.2r_{500}^{\rm H.E.})=S_{\rm X}(0.2r_{500}^{\rm H.E.})$ for the
reason detailed in e.g. Zhang et al. (2007).

The global properties of R1504 obey the scaling relations of nearby
clusters. R1504, for instance, closely matches the $M-Y_{\rm X}$,
$M-M_{\rm gas}$, $M-T$, and $M-L^{\rm cocc}$ relations provided by
e.g. Arnaud et al. (2005, 2007), Vikhlinin et al. (2006), and Zhang et
al. (2008). This supports the finding that the ICM in R1504 is quite
relaxed within $r^{\rm H.E.}_{500}$. Regardless of whether the
high velocity substructure is excluded, R1504 is well within the 26\%
scatter of the $L^{\rm cocc}-\sigma$ relation for the clusters at
similar redshifts studied by Zhang et al. (2011a). According to the
dynamical mass estimate $M^{\rm caustic}_{200}$ and the richness
within $r^{\rm caustic}_{200}$, R1504 agrees with the mass versus
richness scaling relations of Johnston et al. (2007) and Hansen et
al. (2009), who instead used weak-lensing masses.

\subsection{Cluster core}
\label{s:core}

\begin{figure*}
\begin{center}
  \includegraphics[angle=270,width=8cm]{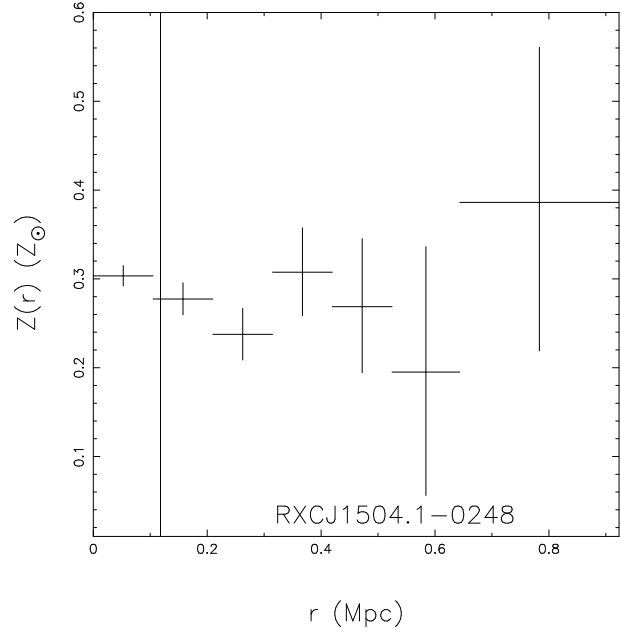}
\end{center}
\caption{Radial distribution of the iron abundance. The vertical line
  denotes $0.1r^{\rm H.E.}_{500}$
\label{f:abun}}
\end{figure*}

\begin{figure*}
\begin{center}
\includegraphics[angle=270,width=8cm]{plots/f7a.ps}
\includegraphics[angle=270,width=8cm]{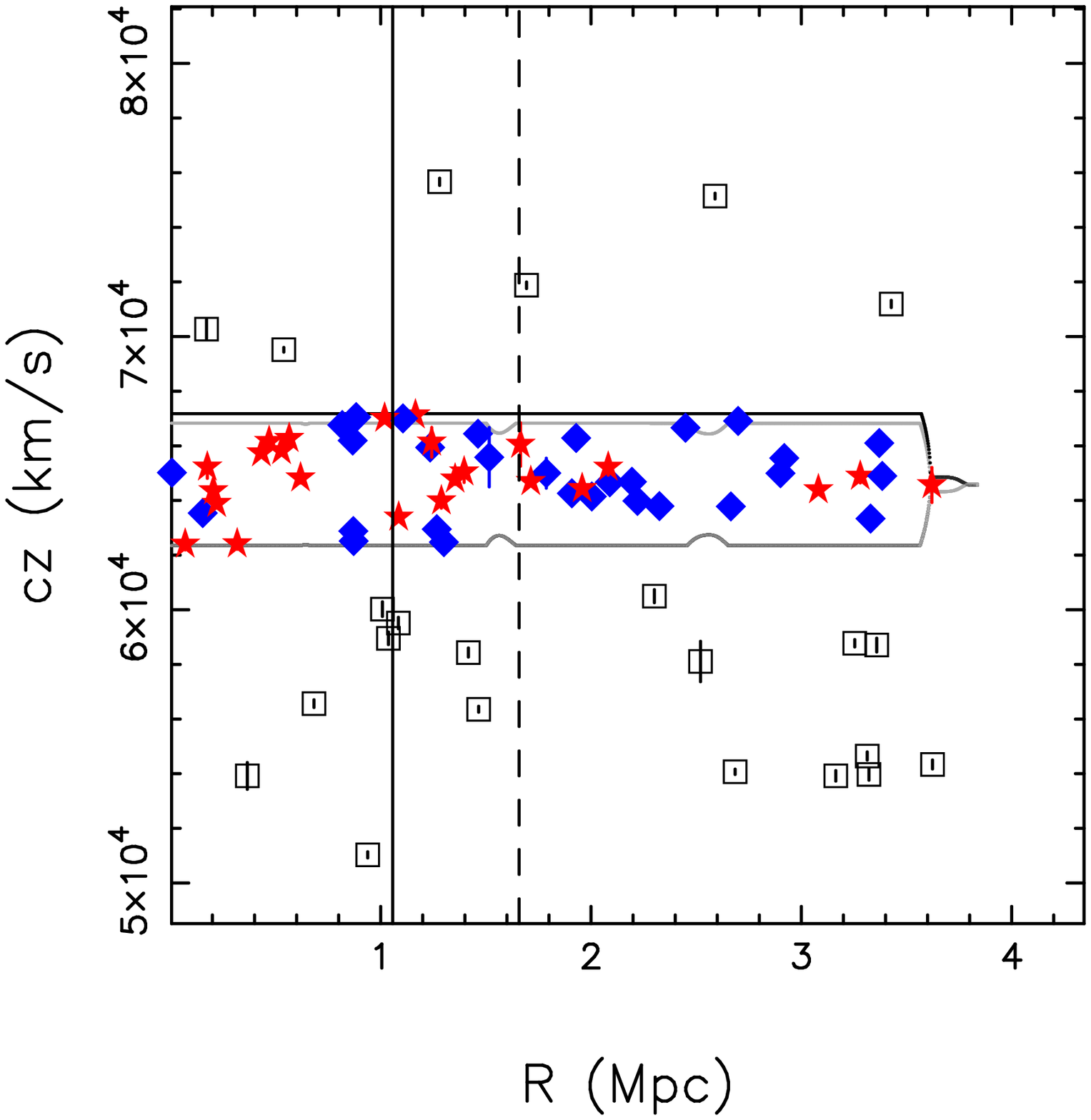}
\end{center}
\caption{{\it Left:} Histogram of the line-of-sight velocities. Member
  galaxies of the cluster, member galaxies of the high velocity group,
  and non-member galaxies are in solid (black), dashed (red), and
  dash-dotted (grey) lines. The arrow shows the cluster redshift. {\it
    Right:} Line-of-sight velocity versus projected radius, in which
  filled blue diamonds and red stars show blue and red member galaxies
  in the caustic (black curve), and open squares show fore- and
  background galaxies. The grey curve shows the symmetric boundary by
  choosing the minimum of the values on both sides of the caustic
  boundary relative to the cluster redshift. The cluster radii,
  $r_{500}^{\rm caustic}$ and $r_{200}^{\rm caustic}$, derived from 
  the mass distribution
  using the caustic method are shown as solid and
  dashed vertical lines, respectively.
  \label{f:caustic}}
\end{figure*}

\begin{figure*}
\begin{center}
\includegraphics[angle=0,width=18cm]{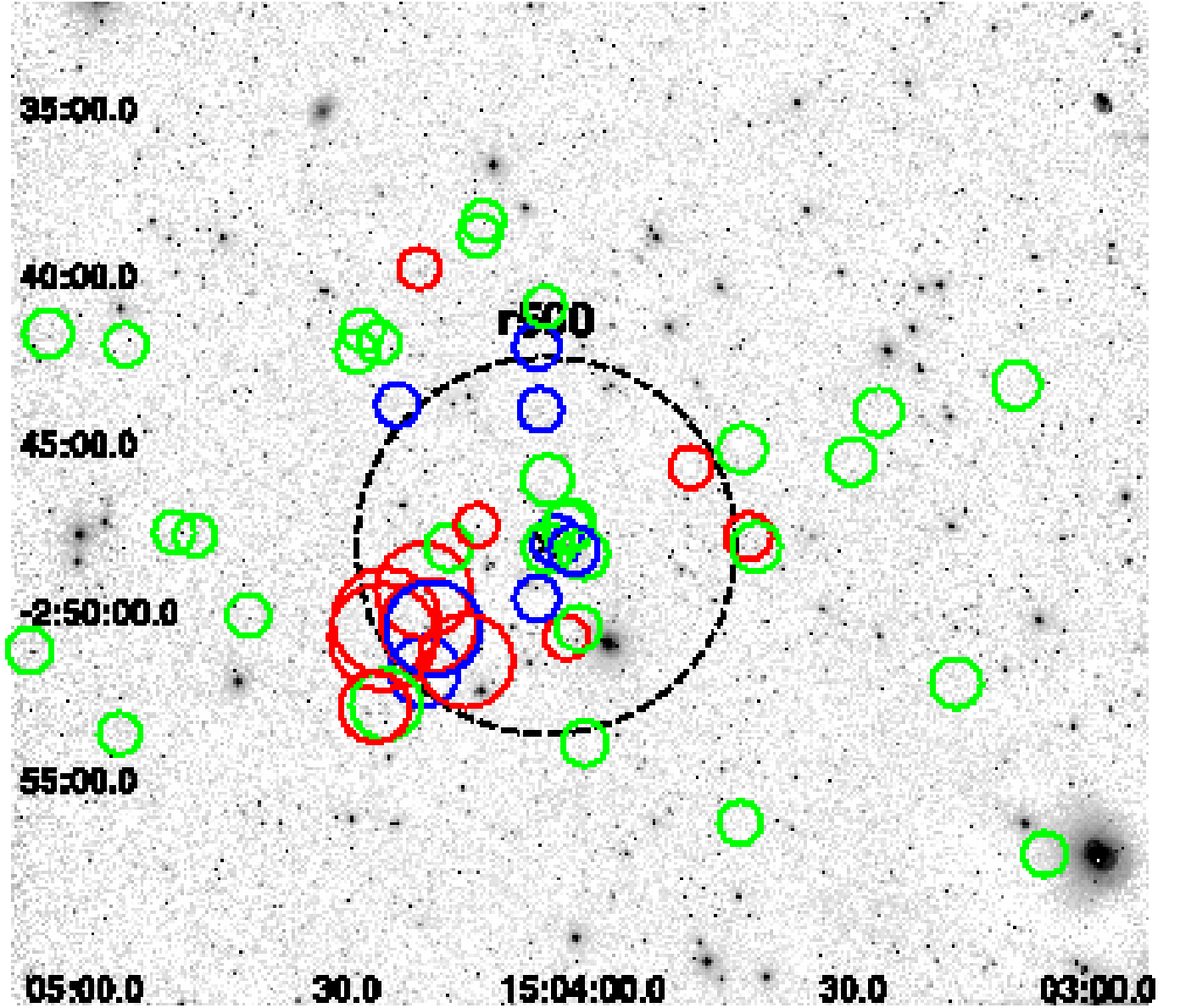}
\end{center}
\caption{WFI $B$-band imaging. 
  Overlaid circles in blue, red and green denote cluster galaxies with 
  spectroscopic follow-up data that have clustercentric 
  velocities toward the observer greater than
  1000\,km\,s$^{-1}$, away from the observer greater than
  1000\,km\,s$^{-1}$, and smaller than 1000\,km\,s$^{-1}$,
  respectively. The radii of these circles are proportional to
    $\exp [\rho_{\rm DS}^2]$. The black cross denotes
  the X-ray flux-weighted centroid and the black dashed circle denotes
  $r^{\rm H.E.}_{500}$. 
  \label{f:wfi}}
\end{figure*}

\begin{figure*}
\begin{center}
\includegraphics[angle=0,width=8.5cm]{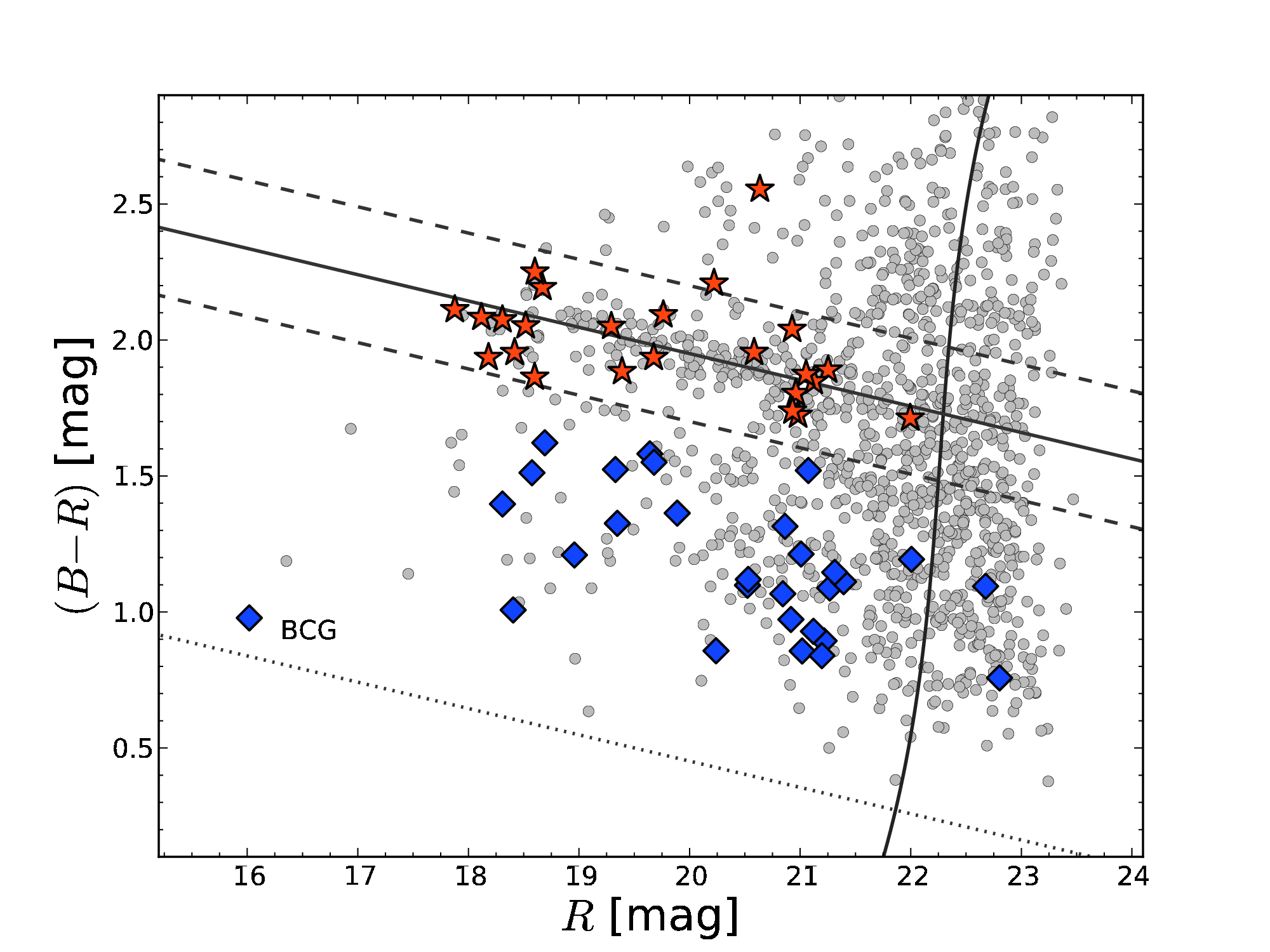}
\includegraphics[angle=0,width=8.5cm]{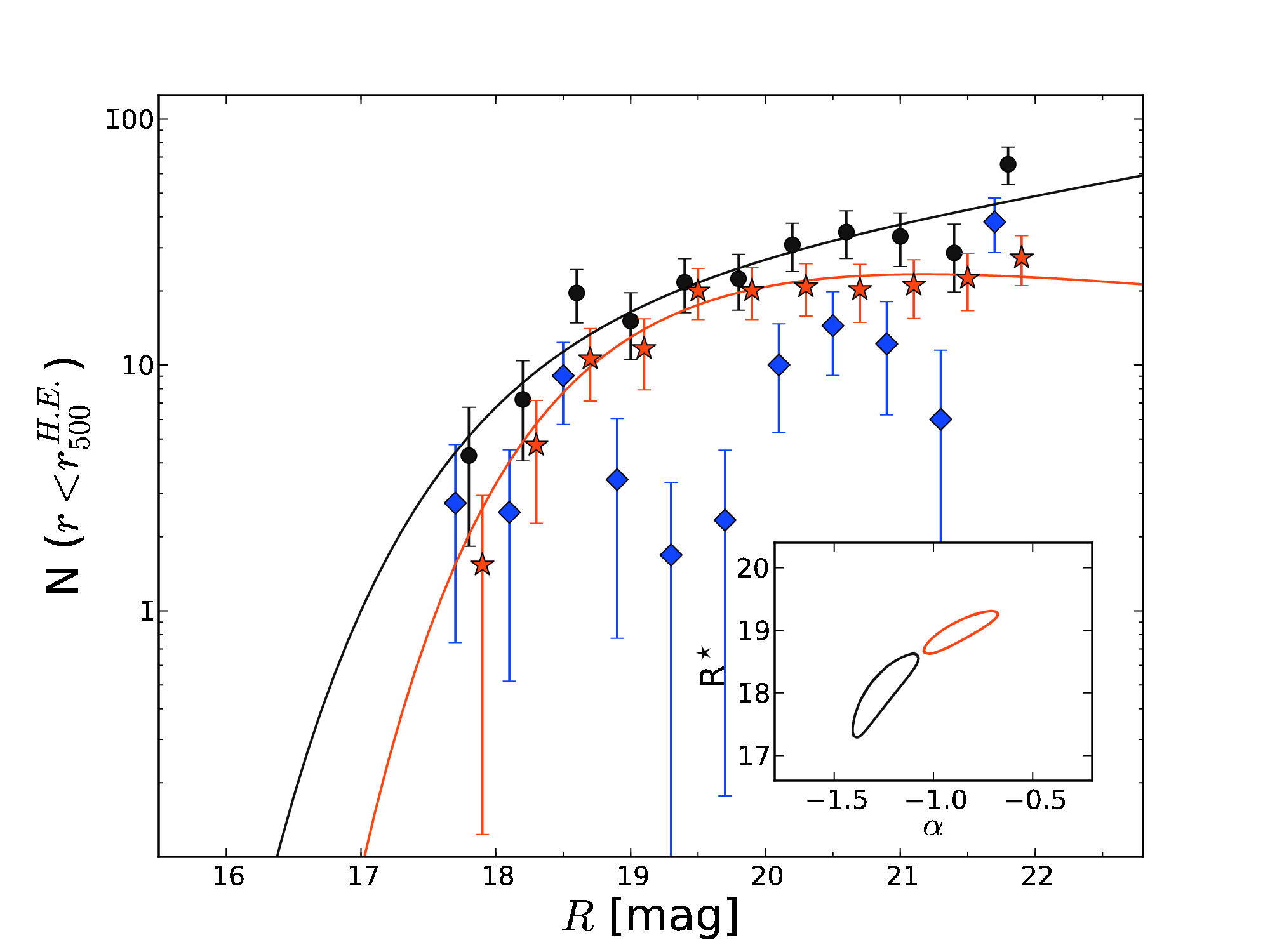}
\end{center}
\caption{ {\it Left:} Colour-magnitude diagram with the corresponding
  red sequence indicated by a solid line and its 
  $3\sigma=0.249$\,magnitude intervals by
  dashed lines. The sample is complete down to the curve at $\sim
  22$~magnitude, which varies slightly with the $B-R$ colour. We thus
  limited the analysis to the galaxies brighter than $R=22$\,magnitude. The
  dotted line denotes the colour cut that is 1.5 magnitude bluer than
  the red sequence, below which the objects were considered as
  foreground contamination. Red stars and blue diamonds indicate red
  and blue spectroscopic member galaxies regardless of their 
  clustercentric distances. The BCG is extremely blue with $B-R=0.98$. {\it
    Right:} Cluster galaxy luminosity functions for red (red stars),
  blue (blue diamonds), and all (black circles) members within $r^{\rm
    H.E.}_{500}$ derived from the X-ray hydrostatic mass estimate. The
  membership is statistically determined using control fields with the
  background contamination statistically subtracted. The best-fit
  Schechter (1976) functions are shown in red and black curves for the
  red and all members, respectively. The inset shows the 68\%
  confidence region for the $R^\star$ and $\alpha$ parameters of the
  best-fit Schechter (1976) functions. No fit was possible for the blue
  members.
  \label{f:rslf}}
\end{figure*}

\begin{figure*}
\begin{center}
\includegraphics[angle=0,width=8.5cm]{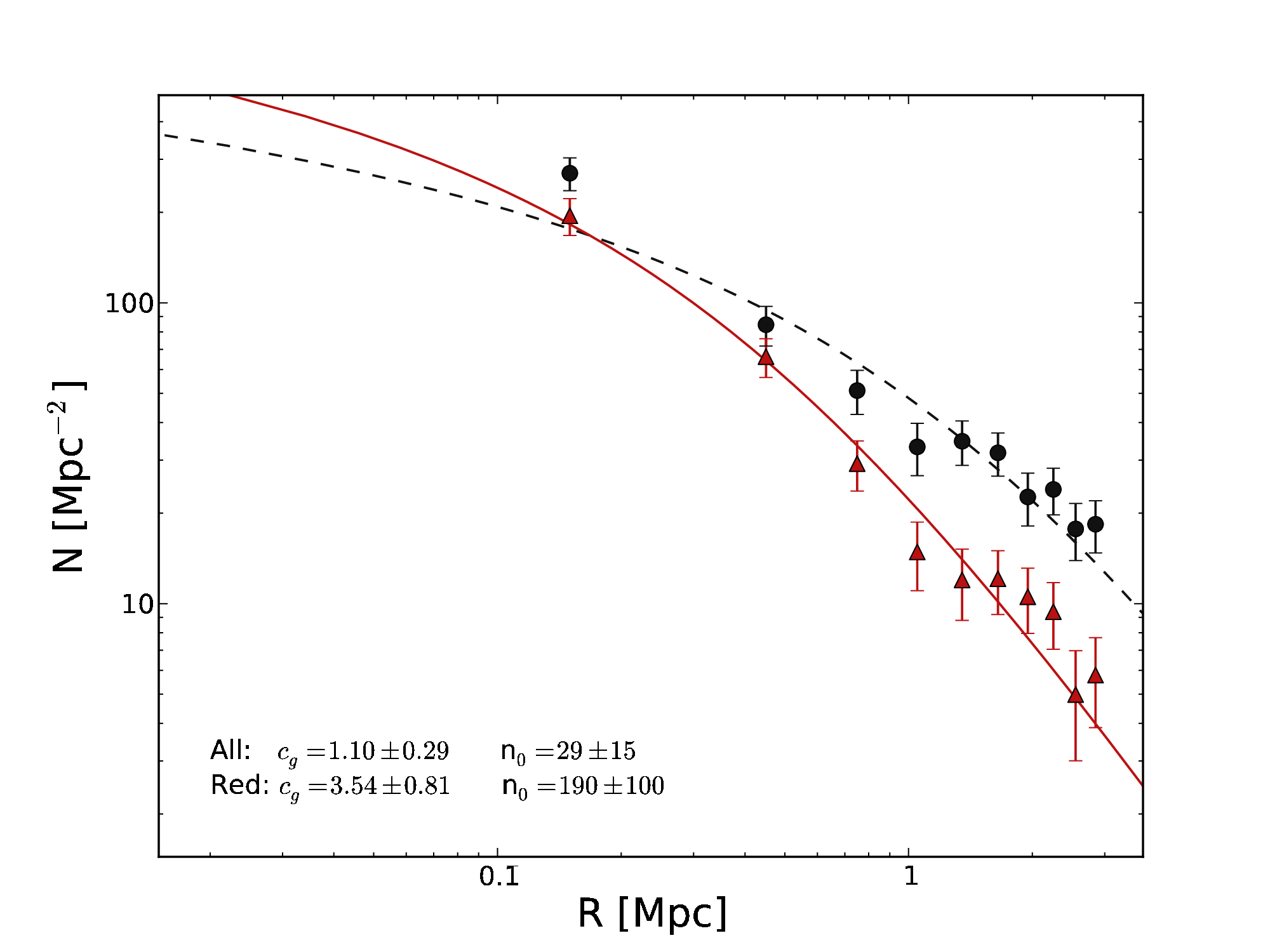}
\includegraphics[angle=0,width=8.5cm]{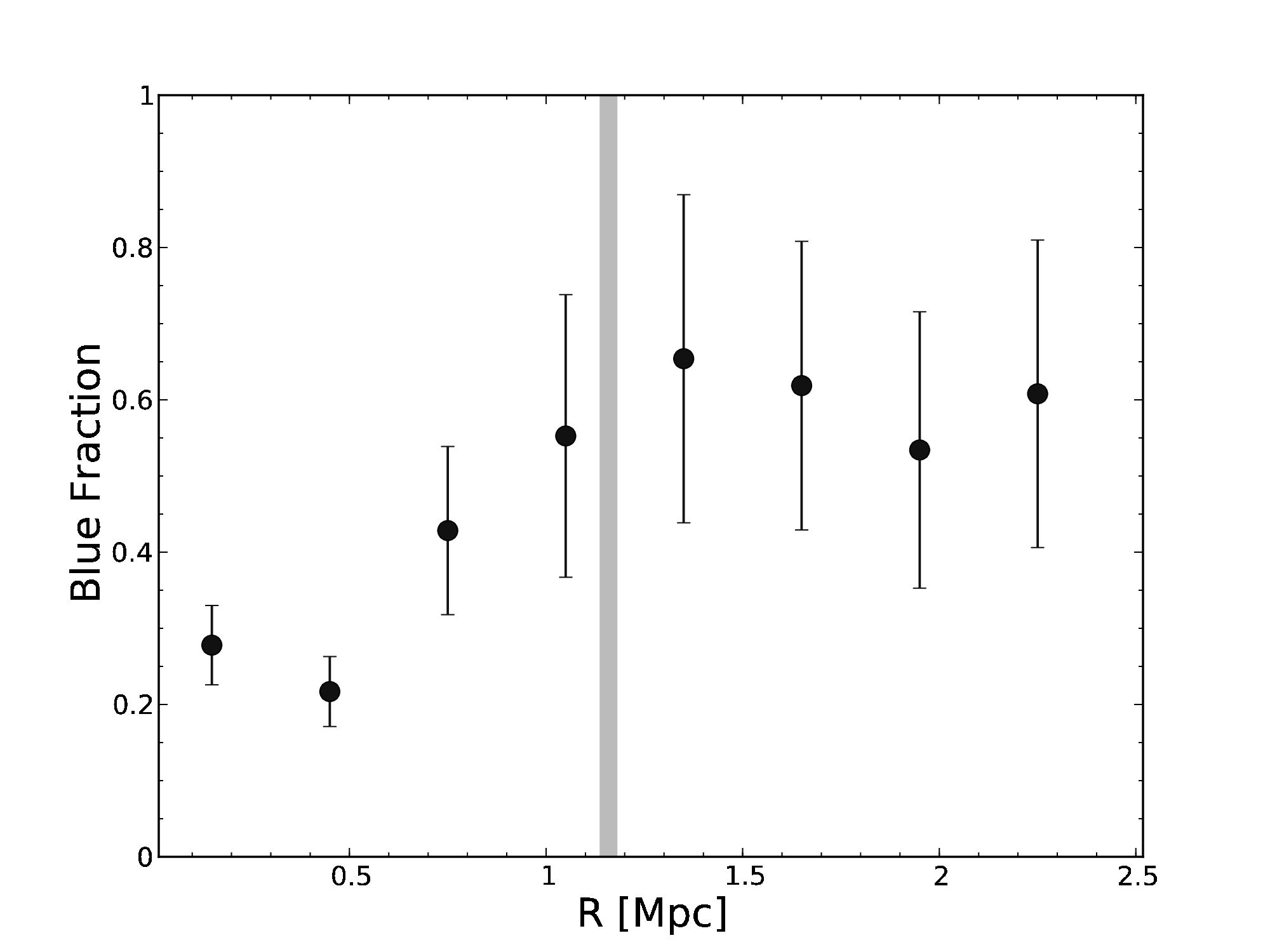}
\end{center}
\caption{{\it Left:} Number density profiles with their best-fit NFW
  models for red member galaxies (red triangles, solid curve) and all
  members (black circles, dashed curve), respectively. We obtained
  reduced $\chi^2=1.03, 2.24$ for the red and all member galaxies,
  which indicates that the red galaxies follow well the theoretical
  profile and the blue galaxies may account for the deviation of the
  fit for all member galaxies. The bin size is 300\,kpc. {\it Right:}
  Ratio of blue to all member galaxies as a function of projection
  distance from the X-ray flux-weighted centroid. The vertical line
  remarks $r^{\rm H.E.}_{500}$.
  \label{f:wfiprof}}
\end{figure*}

\begin{figure*}
\begin{center}
\includegraphics[angle=0,width=8.5cm]{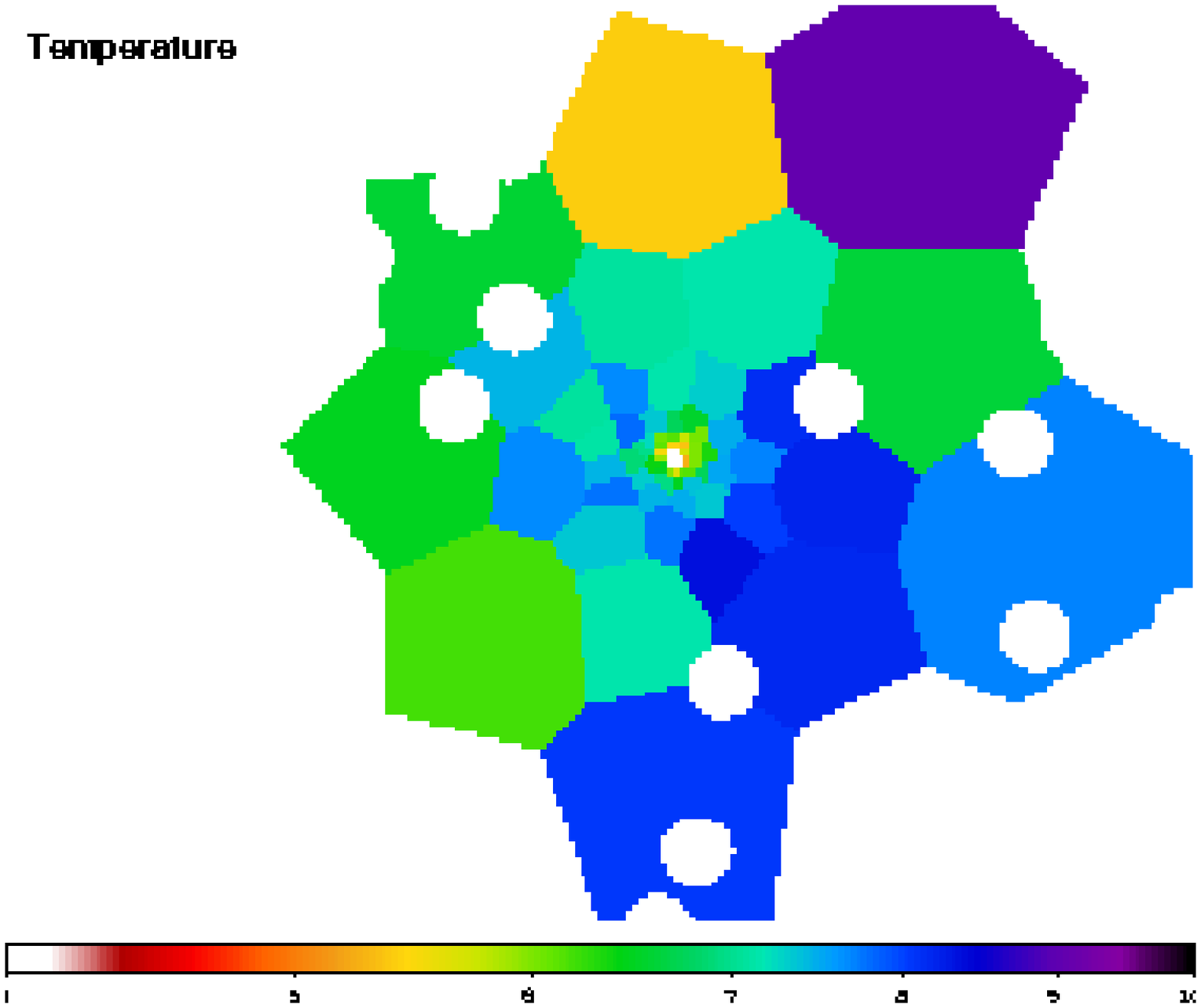}
\includegraphics[angle=0,width=8.5cm]{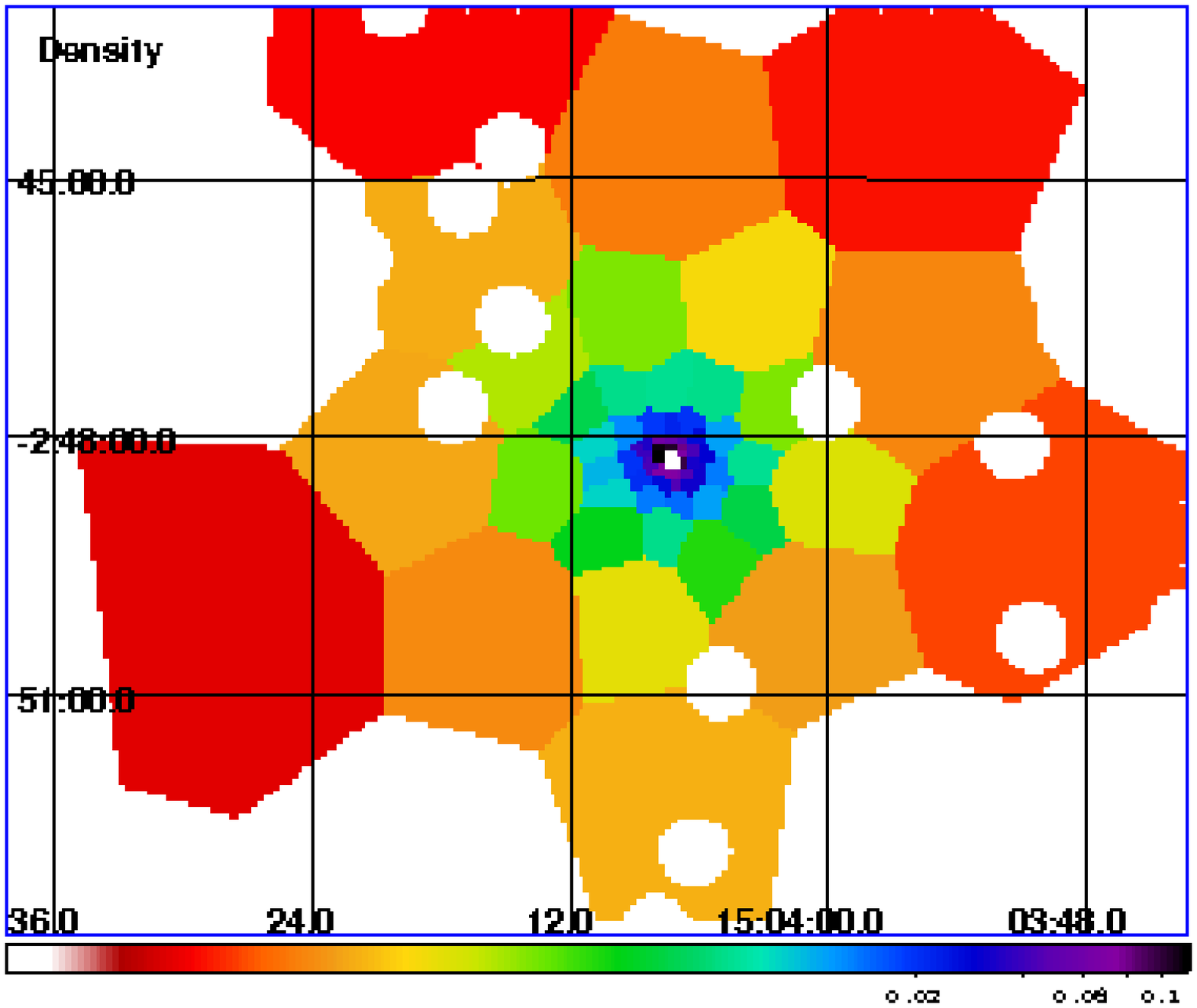}

\includegraphics[angle=0,width=8.5cm]{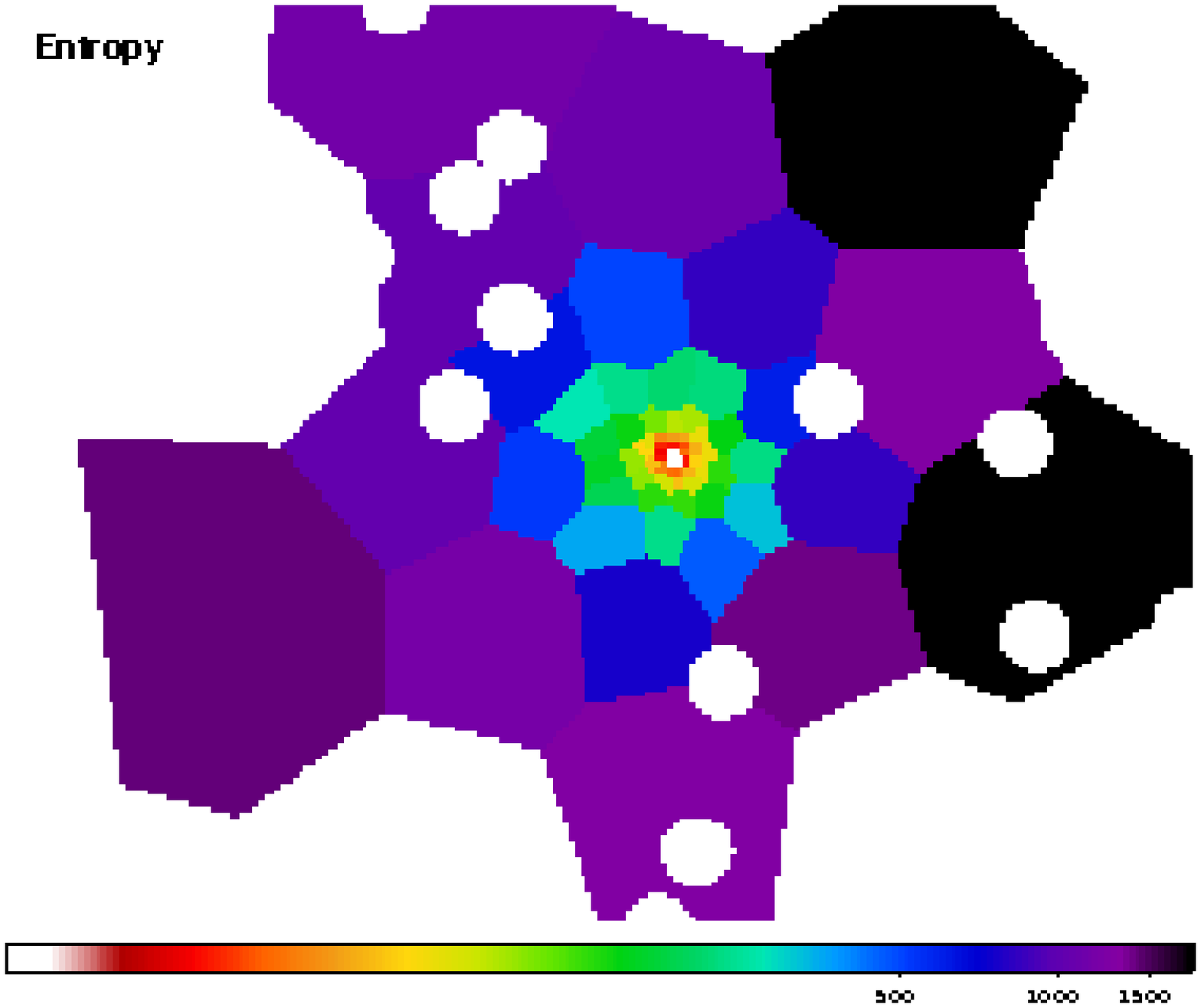}
\includegraphics[angle=0,width=8.5cm]{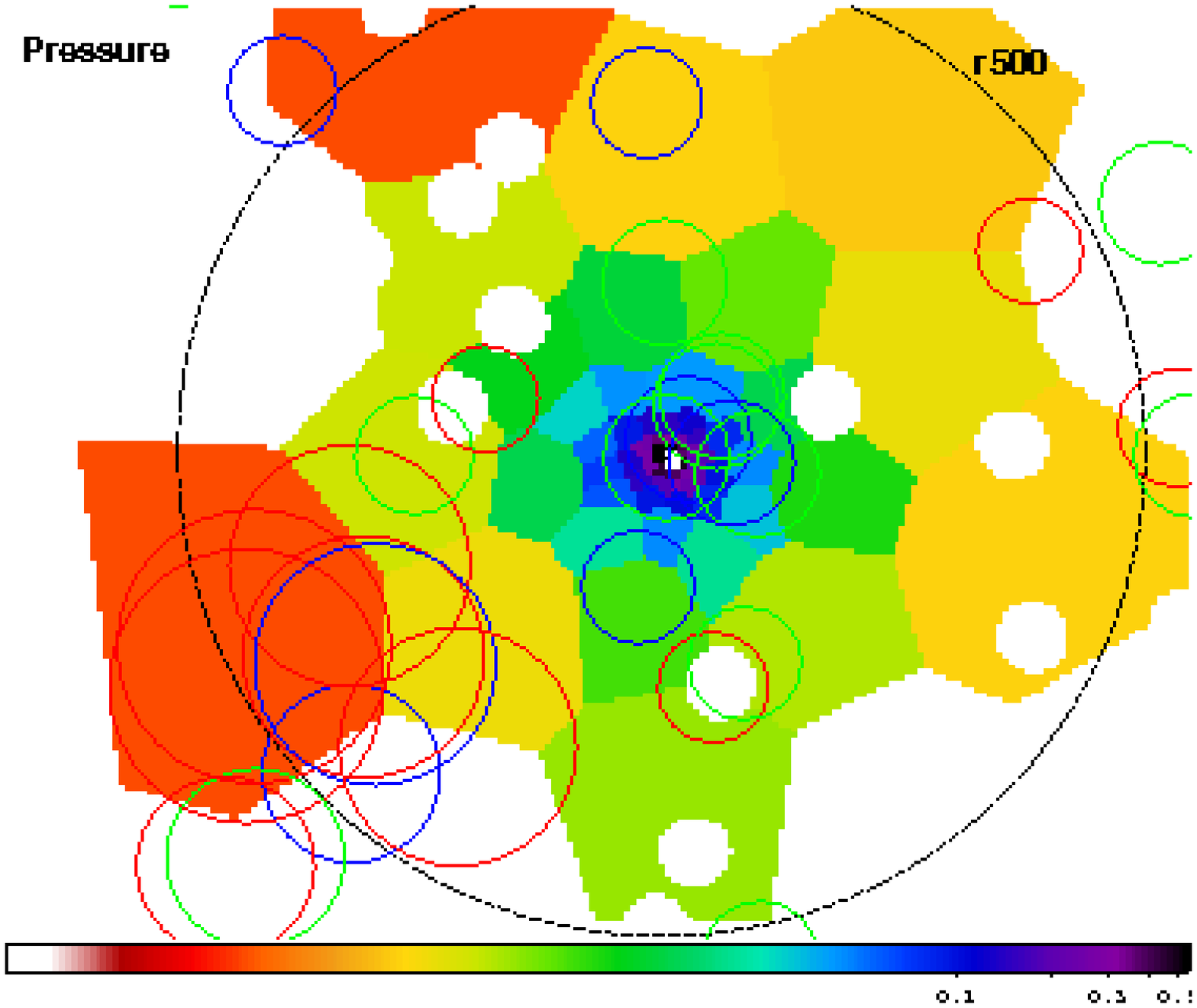}
\end{center}
\caption{Two-dimensional X-ray spectrally measured temperature (upper
  left), electron number-density (upper right), entropy (lower left),
  and pressure (lower right) maps using the Cappellari \& Copin (2003)
  binning scheme. Overlaid circles in blue, red, and green denote
    cluster galaxies that have spectroscopic follow-up data with their
    clustercentric velocities toward the observer greater than
    1000\,km\,s$^{-1}$, away from the observer greater than
    1000\,km\,s$^{-1}$, and smaller than 1000\,km\,s$^{-1}$,
    respectively. The radii of these circles are proportional to
  $\exp [\rho_{\rm DS}^2]$. The black cross denotes the X-ray
  flux-weighted centroid and the black dashed circle denotes $r^{\rm
    H.E.}_{500}$. The white holes mask X-ray detected point
  sources. The colour scales logarithmically from 4 to 10 in keV for
  temperature, from 0.00005 to 0.12 in cm$^{-3}$ for electron number
  density, from 10 to 1800 in keV\,cm$^2$ for entropy, and from 0.0001
  to 0.54 in keV\,cm$^{-3}$ for pressure. \label{f:2dv}}
\end{figure*}

Strong cool-core systems usually show peaked iron abundances in the
cluster cores. However, R1504 displays a relatively flat radial
distribution of iron abundance ranging within [0.2, 0.4] times solar
abundance (Fig.\,\ref{f:abun}). Guo \& Mathews (2010) pointed out in
simulations that AGN outbursts efficiently mix metals in the cluster
core and may even remove the central abundance peak if it is
insufficiently broad.

R1504 is one of nine known clusters hosting radio mini-halos. The
presence of mini-halos supports the above interpretation of mixing
metals in the cluster core. The BCG is blue with a $B-R$ colour of
0.98, and has strong emission lines. The BCG sky position,
RA=15\hh04\mm07\fs573, $\delta$=$-$02\degree48\arcmin14\farcs26
(J2000), is offset by 4\arcsec\ (14~kpc) from the X-ray flux-weighted
centroid, which is within the \emph{XMM-Newton} spacial resolution
with a 6\arcsec\ full width at half maximum. The BCG harbours a radio
source with a brightness of 62\,mJy at 1.4\,GHz (Bauer et
al. 2000). Giacintucci et al. (2011) reveal gas sloshing in the
cluster core, where turbulence may generate particle acceleration to
form mini-halos. The gas sloshing in the cluster core may partially
account for the slightly less-concentrated mass distribution using the
X-ray hydrostatic method than that determined from the caustic method
(Fig.\,\ref{f:mprof}).

\section{Conclusions}
\label{s:con}

We have studied one of the most X-ray luminous clusters of galaxies in
the REFLEX survey, R1504, using \emph{XMM-Newton} imaging
spectroscopy, VLT/VIMOS spectroscopy, and WFI photometry.

The mass distribution determined using the hydrostatic method based on
X-ray imaging spectroscopy agrees with that using the caustic method
based on optical spectroscopy within the $1\sigma$ uncertainties at
most radii, although the former appears to be less concentrated than
the latter. The mass estimate obtained using the caustic method is
80\% of the hydrostatic mass estimate at the X-ray mass determined
radius $r^{\rm H.E.}_{500}$.

At the overdensity of 500, the dynamical mass estimate is 72\% of the
hydrostatic mass estimate. At the overdensity of 200, the
richness-derived mass estimates according to more recent calibrations
of the mass--richness relation (e.g. Johnston et al. 2007; Hansen et
al. 2009) are about 0.8--1.2 times the hydrostatic mass measurement
and 1.0--1.4 times the dynamical mass measurement.

On the basis of the line-of-sight velocities of spectroscopic members,
our DS test has revealed a relatively high-velocity
($>$1000\,km\,s$^{-1}$) group at a projected distance of $\sim r^{\rm
  H.E.}_{500}$ south-east of the cluster centroid. The high velocity
group was also present in the 2-D X-ray maps and weak-lensing shear
map. The dynamical mass estimate is reduced by 14\% when the
substructure of the high velocity group is excluded. This highlights
the importance of a proper analysis of the structures of evolving
systems such as cluster of galaxies in order to obtain precise mass
measurements.

The cluster R1504 has a rather low stellar-mass fraction among nearby
clusters (e.g. Biviano \& Salucci 2006), although there is an
agreement given the large scatter for nearby clusters (e.g. Andreon
2010; Zhang et al. 2011b). Both the intra-cluster light and BCG play
limited roles in accounting for the low stellar-mass fraction.

Within a projected distance of 700\,kpc, the galaxy population is
dominated by red galaxies ($>$70\%) based on photometric data. Beyond
700\,kpc, there is a rapid increase in the fraction of blue
galaxies. The blue-galaxy fraction reaches about 60\% at $r^{\rm
  H.E.}_{500}$ and stays almost constant beyond. The dynamical mass
measurement calculated using blue spectroscopic members is 1.23 times
the value derived using red spectroscopic galaxies.

Despite the high velocity group in the cluster outskirts and the gas
sloshing in the cluster core, R1504 appears to be relatively relaxed
out to $r^{\rm H.E.}_{500}$, and obeys the observed scaling relations
found for similar nearby clusters.

\begin{acknowledgements}
  We acknowledge our referee Andrea Biviano, who provided insight and
  expertise that greatly improved the work. Y.Y.Z. acknowledges Italo
  Balestra, Hans B\"ohringer, Angela Bongiorno, Paola Popesso, David
  Wilman, and Felicia Ziparo for useful
  discussions. Y.Y.Z. acknowledges support by the German BMBF through
  the Verbundforschung under grant 50\,OR\,1103. The \emph{XMM-Newton}
  project is an ESA Science Mission with instruments and contributions
  directly funded by ESA Member States and the USA (NASA). The
  \emph{XMM-Newton} project is supported by the Bundesministerium
  f\"ur Wirtschaft und Technologie/Deutsches Zentrum f\"ur Luft- und
  Raumfahrt (BMWI/DLR, FKZ 50 OX 0001) and the Max-Planck Society. We
  have made use of VLT/VIMOS observations taken with the ESO Telescope
  at the Paranal Observatory under programme 077.A-0058 and WFI
  observations partially supported by the Deutsche
  Forschungsgemeinschaft (DFG) through Transregional Collaborative
  Research Centre TRR 33. The VLT/VIMOS data presented in this paper
  were reduced using the VIMOS Interactive Pipeline and Graphical
  Interface ({\tt VIPGI}) designed by the VIRMOS Consortium. This
  research has made use of the SIMBAD database, operated at CDS,
  Strasbourg, France.
\end{acknowledgements}

\bibliography{Abell}


\newpage 

\begin{table*}
\begin{center}
\caption{Basic properties of R1504.}
\begin{tabular}{ll}
  \hline
  \hline
  \multicolumn{2}{l}{X-ray properties}\\
  \hline
  $r^{\rm H.E.}_{500}$  & $(1.18\pm 0.03)$~Mpc \\
  $M^{\rm H.E.}_{500}$  & $(5.81\pm 0.49)\times 10^{14}M_{\odot}$ \\
  $r^{\rm H.E.}_{200}$  & $(1.76\pm 0.05)$~Mpc \\
  $M^{\rm H.E.}_{200}$  & $(7.65\pm 0.65)\times 10^{14}M_{\odot}$ \\
  $c_{500}$                             & $2.93\pm 0.14$ \\
  $M_{\rm gas,500}(\le r^{\rm H.E.}_{500})$    & $(0.907\pm 0.051)\times 10^{14}M_{\odot}$ \\
  $f_{\rm gas,500}(\le r^{\rm H.E.}_{500})$    & $0.156\pm 0.015$\\ 
  $T_{(0.2-0.5)r^{\rm H.E.}_{500}}$   & ($7.25\pm 0.13$)\,keV\\
  $Z_{(0.2-0.5)r^{\rm H.E.}_{500}}$   & $(0.28\pm 0.03)\;Z_{\odot}$\\
  $L^{\rm incc}_{\rm 0.1-2.4keV,500}(\le r^{\rm H.E.}_{500})$ & $(2.13\pm 0.04)\times 10^{45}$\,erg\,s$^{-1}$\\
  $L^{\rm cocc}_{\rm 0.1-2.4keV,500}(\le r^{\rm H.E.}_{500})$ & $(0.61\pm 0.04)\times 10^{45}$\,erg\,s$^{-1}$\\
  $L^{\rm incc}_{\rm bol,500}(\le r^{\rm H.E.}_{500})$ & $(5.48\pm 0.11)\times 10^{45}$\,erg\,s$^{-1}$\\
  $L^{\rm cocc}_{\rm bol,500}(\le r^{\rm H.E.}_{500})$ & $(1.56\pm 0.11)\times 10^{45}$\,erg\,s$^{-1}$\\
  \hline
  \hline
  \multicolumn{2}{l}{Optical spectroscopic properties with all 53 members}\\
  \hline
  $r^{\rm caustic}_{500}$  & $(1.06\pm0.04)$~Mpc \\
  $M^{\rm caustic}_{500}$  & $(4.17\pm 0.42)\times 10^{14}M_{\odot}$ \\
  $r^{\rm caustic}_{200}$  & $(1.66\pm0.06)$~Mpc \\
  $M^{\rm caustic}_{200}$  & $(6.43\pm 0.65)\times 10^{14}M_{\odot}$ \\
  $M^{\rm caustic}(\le r^{\rm H.E.}_{500})$ & $(4.66\pm 0.47)\times 10^{14}M_{\odot}$ \\
  $\sigma_{\rm 53m}$     & ($1132 \pm 94$)\,km\,s$^{-1}$\\
  $r^{\rm B06}_{\rm 500,53m}$& $(1.217\pm 0.109)$~Mpc\\
  $M^{\rm B06}_{\rm 500,53m}$& $(6.355\pm 1.711)\times 10^{14} M_{\odot}$\\
  \hline
  \hline
  \multicolumn{2}{l}{Optical spectroscopic properties with 48 members excluding the high velocity group}\\
  \hline
  $\sigma_{\rm 48m}$     & ($1079 \pm 97$)\,km\,s$^{-1}$\\
  $r^{\rm B06}_{\rm 500,48m}$& $(1.157\pm 0.111)$~Mpc\\
  $M^{\rm B06}_{\rm 500,48m}$& $(5.452\pm 1.573)\times 10^{14} M_{\odot}$\\
  \hline
  \hline
  \multicolumn{2}{l}{Optical spectroscopic properties with all 29 blue members}\\
  \hline
  $\sigma_{\rm blue}$     & ($1170 \pm 125$)\,km\,s$^{-1}$\\
  $r^{\rm B06}_{\rm 500,blue}$& $(1.260\pm0.141)$~Mpc\\
  $M^{\rm B06}_{\rm 500,blue}$& $(7.057\pm 2.373)\times 10^{14} M_{\odot}$\\
  \hline
  \hline
  \multicolumn{2}{l}{Optical spectroscopic properties with all 24 red members}\\
  \hline
  $\sigma_{\rm red}$     & ($1096 \pm 156$)\,km\,s$^{-1}$\\
  $r^{\rm B06}_{\rm 500,red}$& $(1.175\pm 0.172)$~Mpc\\
  $M^{\rm B06}_{\rm 500,red}$& $(5.722\pm 2.517)\times 10^{14} M_{\odot}$\\
  \hline
  \hline
  \multicolumn{2}{l}{Photometric properties within X-ray determined $r^{\rm H.E.}_{500}$ down to $R=22$\,magnitude}\\
  \hline
  \multirow{2}{*}{Galaxy luminosity function} &
  all members: $R^{\ast}=18.11\pm 0.61$, ~~~$\alpha=-1.24 \pm 0.16$, ~~~$\Phi^{\ast}=20.95 \pm 12.24$\\
  & red members: $R^{\ast}_{\rm red}=18.98\pm 0.34$, $\alpha_{\rm red}=-0.86\pm 0.19$, $\Phi^{\ast}_{\rm red}=34.87\pm 11.75$\\
  $f_{\rm blue}(\le r^{\rm H.E.}_{500})$    & $0.37 \pm 0.06$\\
  $N_{\rm red}(\le r^{\rm H.E.}_{500})$     & $180.2 \pm 15.5$\\
  $N_{\rm all}(\le r^{\rm H.E.}_{500})$     & $283.9 \pm 22.4$\\
  $L_{\rm R,blue}(\le r^{\rm H.E.}_{500})$    & $(0.81 \pm 0.14) \times 10^{12}L_{\odot}$\\
  $L_{\rm R,red}(\le r^{\rm H.E.}_{500})$     & $(1.77 \pm 0.19) \times 10^{12}L_{\odot}$\\
  $L_{\rm R,all}(\le r^{\rm H.E.}_{500})$     & $(2.58 \pm 0.24) \times 10^{12}L_{\odot}$\\
  $M_{\ast,\rm blue}(\le r^{\rm H.E.}_{500})$& $(0.31 \pm 0.07) \times 10^{12}M_{\odot}$\\
  $M_{\ast,\rm red}(\le r^{\rm H.E.}_{500})$ & $(3.42 \pm 0.66) \times 10^{12}M_{\odot}$\\
  $M_{\ast,\rm all}(\le r^{\rm H.E.}_{500})$ & $(3.73 \pm 0.66) \times 10^{12}M_{\odot}$\\
  $f_{\ast,500}(\le r^{\rm H.E.}_{500})$    & $0.0064 \pm 0.0013$\\  
  \hline
  \hline
  \multicolumn{2}{l}{Richness-derived mass measurements down to ($R^{\ast}+1$)\,magnitude}\\
  \hline
  $r^{\rm J07}_{200}$     & $(1.851 \pm 0.082)$~Mpc\\
  $M^{\rm J07}_{200}$     & $(8.948 \pm 1.190)\times 10^{14} M_{\odot}$\\
  $N^{J07}_{200}$        & $92.67\pm 9.63$\\
  $r^{\rm H09}_{200}$     & $(1.651 \pm 0.077)$~Mpc\\
  $M^{\rm H09}_{200}$     & $(6.347\pm 0.887)\times 10^{14} M_{\odot}$\\
  $N^{H09}_{200}$        & $83.89\pm 9.16$\\
  \hline
  \hline
\end{tabular} 
\label{t:log}
\end{center}
\hspace*{0.3cm}{\footnotesize }
\end{table*}

\newpage


\clearpage
\appendix

\section{X-ray 2-D spectrally measured maps}
\label{a:2d}

\end{document}